\documentclass[footinbib,twocolumn,showpacs,amsmath,amstex,amssymb,mathfonts,superscriptaddress,prx]{revtex4-1}
\usepackage{graphicx}
\usepackage{color}
\usepackage{bm}
\usepackage{amsmath}
\usepackage{amssymb}
\usepackage{amsthm}
\usepackage{amsfonts}
\usepackage[colorlinks,bookmarks=true,citecolor=blue,linkcolor=red,urlcolor=blue]{hyperref}

\usepackage{bbm}

\newcommand*{\id}{{\normalfont\hbox{1\kern-0.15em \vrule width .8pt depth-.5pt}}}

\newcommand{\be}{\begin{equation}}
\newcommand{\ee}{\end{equation}}

\newcommand{\la}{\label}
\newcommand{\bea}{\begin{eqnarray}}
\newcommand{\eea}{\end{eqnarray}}

\begin{document}

\title{Geometric quench and nonequilibrium dynamics of fractional quantum Hall states}
\author{Zhao Liu}
\affiliation{Zhejiang Institute of Modern Physics, Zhejiang University, Hangzhou 310027, China}
\affiliation{Freie Universit\"at Berlin, Dahlem Center for Complex Quantum Systems and Institut f\"ur Theoretische Physik, Arnimallee 14, 14195 Berlin, Germany}
\author{Andrey Gromov}
\affiliation{Materials Sciences Division, Lawrence Berkeley National Laboratory and Department of Physics, University of California, Berkeley, California 94720, USA}
\affiliation{Kadanoff Center for Theoretical Physics, University of Chicago, Illinois 60637, USA}
\author{Zlatko Papi\'c}
\affiliation{School of Physics and Astronomy, University of Leeds, Leeds LS2 9JT, United Kingdom}
\pacs{73.43.Lp, 71.10.Pm}

\date{\today}
\begin{abstract}
We introduce a quench of the geometry of Landau level orbitals as a probe of nonequilibrium dynamics of fractional quantum Hall (FQH) states. We show that such geometric quenches induce coherent many-body dynamics of neutral degrees of freedom of FQH fluids. The simplest case of mass anisotropy quench can be experimentally implemented as a sudden tilt of the magnetic field, and the resulting dynamics  reduces to the harmonic motion of the spin-2 ``graviton" mode, i.e., the long wavelength limit of the Girvin-MacDonald-Platzman magnetoroton.  We derive an analytical description of  the graviton dynamics using the bimetric theory of FQH states, and find agreement with exact numerical simulations at short times. We show that certain types of geometric quenches excite higher-spin collective modes, thus establishing their existence in a microscopic model and motivating an extension of geometric theories of FQH states. 
\end{abstract}

\maketitle

\section{Introduction}

Fractional quantum Hall (FQH) states are an epitome of topological phases of matter, featuring exotic phenomena such as fractionalization~\cite{Laughlin-PhysRevLett.50.1395}, topological order~\cite{wen1990topological}, and protected edge excitations~\cite{halperin1982quantized, WenEdge}. These phenomena arise as emergent properties of a two-dimensional (2D) electron system in a perpendicular magnetic field \cite{Prange}. As a consequence of Landau level quantization, electrons in these systems have negligible kinetic energy, which paves the way for Coulomb interaction to produce a variety of exotic effects, whose experimental hallmark is the fractional quantization of the Hall conductance \cite{Tsui-PhysRevLett.48.1559}. 

Given the intricate, purely interacting nature of FQH states, it comes as no surprise that the existing research mainly focused on their equilibrium (static) properties at zero temperature. These properties have often been described using two closely related frameworks: trial wave functions~\cite{Laughlin-PhysRevLett.50.1395, Jain:1989p294, Moore1991362, ReadRezayiParafermion} and topological quantum field theory (TQFT)~\cite{zhang1989effective}, which encode the linear response functions~\cite{wen1992shift, gromov2015framing}, exotic quasiparticles  (``anyons")~\cite{LeinaasMyrheim, WilczekAnyons} characterized by the fractional charge, spin and statistics~\cite{wen1992classification}, robust edge modes~\cite{WenEdge} and the topological degeneracy on a torus \cite{haldane1985periodic, wen1990topological}. 
The predictions of microscopic trial states and TQFT have been tested in numerical simulations of unprecedented accuracy, in particular in the case of Laughlin~\cite{Laughlin-PhysRevLett.50.1395, Haldane-Rezayi-ED}  and composite fermion states~\cite{jainbook}.

Besides the low-energy properties, the complete understanding of a quantum system  requires the understanding of its \emph{dynamics}, which is determined by the system's excited states (possibly at non-zero energy density). This dynamics can be physically probed using the global quantum quench: prepare the system in its ground state $|\psi_0\rangle$;  abruptly change the Hamiltonian, $H\to H'$; let the system evolve and perform measurements on the time-evolved state. For systems that are intrinsically decoupled from the environment, this quench amounts to the innocuous-looking Schr\"odinger evolution, $|\psi(t)\rangle = \exp(-i H' t)|\psi_0\rangle$, where  $|\psi_0\rangle$, crucially, is \emph{not} an eigenstate of the quench Hamiltonian, $H'$. 
Even though the Schr\"odinger evolution can be written in such a compact way, it can nevertheless produce  incredibly complex outcomes as it might involve highly excited eigenstates of $H'$. In one-dimensional integrable systems, where such eigenstates can be computed using techniques like algebraic Bethe ansatz, a particularly deep theoretical understanding of quenches has been achieved (see the recent review~\cite{EsslerReview}) and verified in cold atom experiments~\cite{Kinoshita2006}. Other notable examples include one-dimensional critical systems where conformal invariance allows one to analytically describe the post-quench dynamics of various observables (see Ref.~\cite{CalabreseCardy} and references therein) and their holographic duals \cite{nozaki2013holographic}, where quantum quench is interpreted as a process of black hole formation and decay.

In this work, we introduce a new type of \emph{geometric quench} to study the dynamics of FQH states out of equilibrium. While in integrable systems the term ``geometric quench" refers to a sudden change of the size of system~\cite{GeometricQuenchMeisner, GeometricQuenchBuljan, GeometricQuenchCaux, GeometricQuenchAlba} (e.g., controlled by the trap potential in cold atomic systems), our setup assumes that the size of the system remains fixed, but the geometry of Landau level orbitals undergoes an abrupt change. This way of probing the system is quite physical: changing the geometry of Landau level orbitals can be achieved directly by tilting the magnetic field, i.e., by introducing a component of magnetic field tangential to the plane occupied by the particles. The tilted field technique is regularly employed in FQH experiments to measure the spin polarization of various states~\cite{EisensteinTilt}, while more recently it has also  been used to map out the anisotropy of the composite fermion Fermi surface~\cite{Kamburov}. 

Geometric quench is designed to excite the neutral degrees of freedom of FQH states, which have certain universal features that are geometric in nature. 
Two major examples are the geometric response function known as Hall viscosity \cite{avron1995viscosity, read2009non, HaldaneViscosity} and the geometric degrees of freedom responsible for the nematic transition \cite{maciejko2013field, you2014theory}. The former can be understood in terms of response to the variations of ambient geometry \cite{ferrari2014fqhe, abanov2014electromagnetic, douglas2010bergman, gromov2015framing, BradlynRead, CanLaskinWiegmann, klevtsov2015geometric, hughes2011torsional, gromov2014density}, while the latter can be formulated as fluctuating geometry \cite{HaldaneGeometry, GromovSon} (see also Ref.~\cite{wiegmann2017nonlinear} for a related perspective).
We will show that the real-time dynamics following the geometric quench is accurately described by the excitations of geometric degrees of freedom that characterize the underlying FQH states~\cite{HaldaneGeometry}. 

More specifically, in the case of Laughlin states, we will show that a (adiabatic) geometric quench driven by an abrupt change of the effective mass tensor of particles excites a single degree of freedom that carries angular momentum (or spin) $L=2$, which can be viewed as a fluctuating, ``intrinsic metric''~\cite{HaldaneGeometry2}. This spin-$2$ degree of freedom is described by a symmetric matrix, which makes it formally similar to the fluctuating space-time metric --- a variable that one would like to use in a theory of quantum gravity. For the lack of a better term we will continue to refer to this geometric degree of freedom as ``graviton"~\cite{yang2012model,Golkar2016}. The geometric quench thus allows to excite the emergent FQH graviton and to observe its non-trivial dynamics in time. One has to bear in mind that the graviton discussed in this paper is non-relativistic and massive (i.e., gapped). The closest relativistic cousin of the geometric aspects of the FQH problem is the zwei-dreibein theory of $3$D massive gravity~\cite{bergshoeff2013zwei} (see also Ref.~\cite{bergshoeff2018gravity}).  

In order to model the geometric quench, we utilize two recent advances in the theory of the FQH effect: the formulation of the bimetric theory~\cite{GromovSon,nguyen2017fractional}, which describes topological properties of FQH states on curved surfaces whilst incorporating the mentioned spin-$2$ excitation, and the generalization of the Haldane pseudopotentials~\cite{GeneralizedPPs} to systems with broken rotational invariance. Bootstrapping these methods, we derive an analytical description of the real-time dynamics of FQH states following a geometric quench, and confirm it against numerical simulations for the case of Laughlin states.

Our main findings are the following: (i) we demonstrate that geometric quench induces the dynamics of neutral degrees of freedom in FQH states which have geometric character; (ii) in the simplest type of quenches, which are driven by a mass tensor deformation in the Laughlin phase, we show that the post-quench dynamics is determined by an exponentially small fraction of excited states with quadrupolar order (i.e., spin-$2$) in the continuum of the spectrum, whose energy sets the oscillation frequency; (iii) we generalize the bimetric theory~\cite{GromovSon} to describe a quench, and obtain an analytical description of the dynamics following the quench [Eqs.~(\ref{eq:phi}) and (\ref{eq:Q})]; (iv) we show that this analytical description is in excellent agreement with exact numerical simulations of geometric quench in (ii) for finite systems at short times, and generally captures well the oscillations in the dynamics at moderate times, thus justifying the validity of bimetric theory; (v) we design more complex types of geometric quenches which excite the higher-spin excitations in the spectrum of FQH states.  We note that higher-spin symmetry has attracted much interest in various areas of theoretical physics, including phase space Fermi fluid \cite{dhar1992non}, collective field theory~\cite{jevicki1993quantum}, generalization of gauge/gravity dualities~\cite{giombi2014partition}, dynamics of incompressible, inviscid fluids~\cite{goldin1987diffeomorphism, arnold1992topological}, large $N$ gauge theory~\cite{bars1990strings}, etc. The presence of higher spin excitations in FQH states, here unambiguously identified via their distinct dynamical response, reveals a much richer structure in FQH states, including the simplest Laughlin states, and calls for an extension of their geometric description~\cite{HaldaneGeometry}.

The remainder of the paper is organized as follows. In Sec.~\ref{sec:quench}, we motivate the geometric quench and provide details of its setup.  In Sec.~\ref{sec:numerics}, we introduce our method, and show that generalized pseudopotentials allow us to identify states that contribute to the quench dynamics.
 Section~\ref{sec:results} contains the bulk of our numerical results for the simplest type of quench driven by changing the metric of Landau level orbitals. An analytical description of such quenches is developed in Sec.~\ref{sec:bimetric}  in the framework of the bimetric theory. In Sec.~\ref{sec:higherspin}, we provide an example of a more complex type of quench which excites higher-spin modes in the spectrum, which currently lacks a theoretical description. Our conclusions are presented in Sec.~\ref{sec:conclusions}, while Appendixes contain additional data and extensions of our results to more realistic (non-instantaneous) quenches and lattice models of fractional Chern insulators.

\section{Geometric quench of a fractional quantum Hall state}\label{sec:quench}

The Hamiltonian describing an FQH system is given by~\cite{Prange}
\begin{eqnarray}\label{eq:ham}
H =\frac{1}{N_\Phi}\sum_{\mathbf{q}} \bar V_\mathbf{q} \bar \rho_\mathbf{q} \bar \rho_{-\mathbf{q}},
\end{eqnarray}
where the sum runs over the 2D Brillouin zone (threaded by $N_\Phi$ flux quanta) and $\bar\rho_\mathbf{q} = \sum_{j=1}^{N} e^{i \mathbf{q}\cdot \mathbf{R}_j}$ is the density operator projected to a single Landau level (LL). The projection to a LL imposes the non-commutative geometry of the ``guiding center coordinates," $\left[ R_i^a, R_j^b \right] = -i\ell_B^2\epsilon^{ab}\delta_{i,j}$, where $\ell_B=\sqrt{\hbar/eB}$ is the magnetic length~\cite{Prange}. This non-trivial constraint, along with the absence of kinetic energy in Eq.~(\ref{eq:ham}),  gives rise to a wide variety of ordered phases, depending on the filling factor $\nu=N/N_\Phi$ and the details of the (projected) interaction potential $\bar V_\mathbf{q}$. Note that because the guiding centers do not commute, bosonic FQH states are also possible and realize a similar variety of phases~\cite{CooperWilkinGunn, RegnaultJolicoeur}. In this paper, we focus on the simplest Laughlin states, which occur at filling $\nu=1/2$ for bosons and $\nu=1/3$ for fermions. 

\begin{figure}[t]
\includegraphics[width=\linewidth]{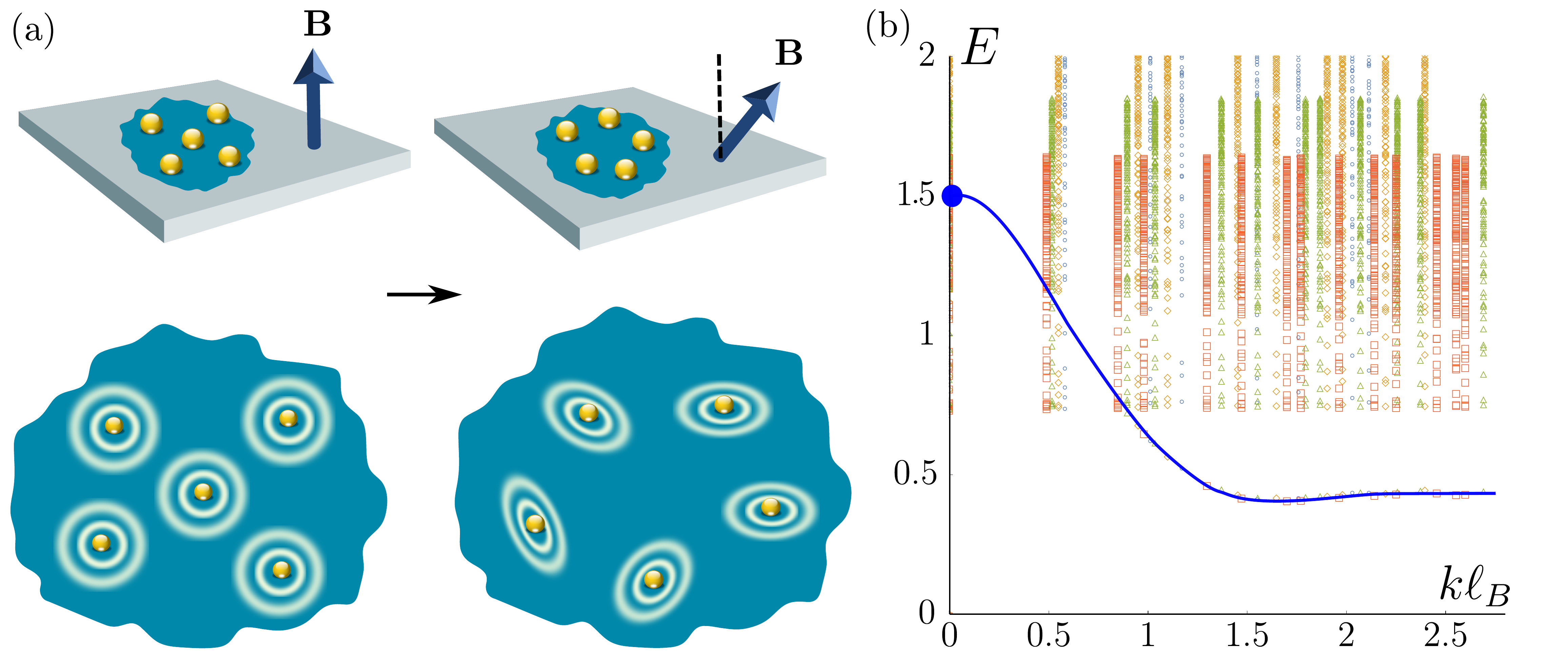}
\caption{(a) Geometric quench protocol: instantaneous tilt of the magnetic field induces dynamics of the intrinsic metric, which describes the shape of particle-flux composites that form an FQH state. (b) The excitation spectrum of the $\nu=1/3$ Laughlin state, obtained by exact diagonalization of its $V_1$ parent Hamiltonian, in systems with $N=7-10$ electrons. Blue line traces out the SMA collective mode~\cite{GMP85}, which transforms into an emergent graviton excitation in the $k\to 0$ limit (around energy $E\approx 1.5$). It will be shown in Sec.~\ref{sec:bimetric} that geometric quench induces oscillatory dynamics of the $k\to 0$ limit of the graviton mode. Note that this mode is formed by excited states in the \emph{continuum} of the energy spectrum.} 
\label{fig:quench}
\end{figure}

Beside the non-commutative constraint, geometry appears in an FQH problem in three distinct guises. First, the interaction potential $\bar V_\mathbf{q} = V_\mathbf{q} |F_m(\mathbf{q})|^2$ is a function of two independent tensors $g_m$ and $g_i$~\cite{HaldaneGeometry}, where $F_m(\mathbf{q})=\exp(-g_m^{ab}q_aq_b\ell_B^2/4)$ is the (lowest) LL form factor and $V_\mathbf{q}$ in general depends on $\sqrt{g_i^{ab}q_a q_b}$ (for the Coulomb interaction, $V_\mathbf{q}=2\pi/\sqrt{g_i^{ab}q_a q_b}$).
These two tensors are imposed by \emph{extrinsic} experimental conditions, such as the direction of the magnetic field and properties of the underlying solid-state material. $g_m$ has physical origin in the band mass tensor and can be conveniently parametrized by a $2\times 2$ unimodular matrix (${\rm det} g_m =1$)
\begin{eqnarray}
g_m = \left(
\begin{array}{cc}
 \cosh Q +\cos \phi  \sinh Q & \sin \phi  \sinh Q \\
 \sin \phi  \sinh Q & \cosh Q-\cos \phi  \sinh Q \\
\end{array}
\right),\;\;\;\;\;\label{eq:ghat0}
\end{eqnarray}
where $Q$ and $\phi$ are real numbers that parametrize a stretch or rotation of the tensor. On the other hand, $g_i$ is in general different from $g_m$. In the case of Coulomb interaction, $g_i$ originates from the dielectric tensor of the material that hosts the FQH system. Note that both $g_m$ and $g_i$ define the shape of a circle: $g_m$ enters the single-particle wave function thus defining the shape of LL orbitals, while $g_i$ determines the shape of the interaction equipotentials. They hence effectively measure the distance from the ``circle center,'' and in this sense we alternatively refer to them as extrinsic metrics~\cite{HaldaneGeometry}.

Subject to these two extrinsic metrics, an FQH state develops a third, \emph{intrinsic} geometric degree of freedom. As a many-body property of the system, this intrinsic degree of freedom defines the shape of particle-flux composite droplets in FQH ground states and can also be thought of as a metric $\hat g$ parametrized by the same matrix as in Eq.~(\ref{eq:ghat0}). If the extrinsic metrics are equal ($g_m=g_i$), but not necessarily isotropic ($\neq \id$), the intrinsic metric is equal to them ($\hat g = g_m = g_i$) and the physical state is isotropic in a transformed coordinate frame. More generally, when $g_m \neq g_i$, $\hat g$ is determined from energetic compromise between $g_m$ and $g_i$, and is in general different from both of them~\cite{BoYangPhysRevB.85.165318, XinWanPhysRevB.86.035122}.  The intrinsic metric, as an emergent many-body property of the system, will be the main focus of the quench dynamics in this paper.

The dynamics of $\hat g$ can in principle be induced by changing the mass tensor $g_m$. While this might be feasible in materials such as AlAs~\cite{Gokmen2010}, a much more practical way is to tilt the magnetic field [Fig.~\ref{fig:quench}(a)]. 
Semiclassically, the particles still prefer to make circular orbits around the tilted direction, but because they are confined to a narrow sample, their orbits deform into ellipses in the plane of the FQH system, hence giving rise to an effective mass anisotropy. In fact, assuming a parabolic confining potential in the perpendicular direction, tilt can be exactly represented as a $2\times 2$ anisotropic mass tensor~\cite{BoYangPhysRevB.85.165318, PapicTilt, BoYangTilt}. Hence, to an excellent approximation, we can model the effect of a tilt by an anisotropic mass tensor like in Eq.~(\ref{eq:ghat0})~\cite{PapicTilt}. 

Our quench protocol can now be defined as follows: (i) prepare a Laughlin state $|\psi_0\rangle$ as the ground state of $H$ with mass tensor $g_m$; (ii) instantaneously change $g_m \to g_m'$; (iii) evolve in time assuming the system is closed, i.e., $|\psi(t)\rangle =\exp(-i H' t) |\psi_0\rangle$, and measure the intrinsic metric as a function of time, $\hat g(t)$.  This protocol is the minimal theoretical model for an experiment where the magnetic field is suddenly tilted. At considerable computational expense, the model can be made more realistic by directly including a parallel field (instead of approximating it by mass anisotropy), relaxing the assumption of a closed system, etc. In later sections of the paper, we will consider some generalizations of the protocol where the quench cannot be described in terms of a $2 \times 2$ metric (Sec.~\ref{sec:higherspin}) or when it is not applied instantaneously (Appendix~\ref{sec:noninst}).

The intuition behind the geometric quench is the following. In a purely perpendicular field, an FQH state is a low-entangled state of composite objects -- particles surrounded by correlation holes of certain size. For example, in the $\nu=1/3$ Laughlin state, we can view them as electrons in an area corresponding to three magnetic flux quanta [Fig.~\ref{fig:quench}(a)]. These objects have a finite area (fixed by the electron density and total flux through the system), but their shape can vary and is determined by the intrinsic metric $\hat g$ of the FQH state~\cite{Johri}, or equivalently by two real parameters, $Q$ and $\phi$, as in Eq.~(\ref{eq:ghat0}). For a adiabatic quench that keeps the system within the Laughlin phase, these droplets are expected to fluctuate and our goal is to determine the equations of motion for $Q$ and $\phi$. This is, however, a non-trivial problem because the representation of a state in Fig.~\ref{fig:quench}(a) is merely a cartoon (e.g., the operators projected onto the droplets do not commute with one another).

As emphasized earlier, the key to understanding dynamics are the excited eigenstates in the spectrum of $H'$. Such a spectrum for the Laughlin $\nu=1/3$ state, obtained by diagonalization of its parent Hamiltonian -- the $V_1$ Haldane pseudopotential~\cite{Haldane1983}-- is shown in Fig.~\ref{fig:quench}(b).  The low-energy spectrum is dominated by a collective mode known as the Girvin-MacDonald-Platzman (GMP) mode or ``magnetoroton"~\cite{GMP85, GMP86}. This mode is accurately described by the single-mode approximation (SMA), and intuitively represents a density modulation on top of the ground state given by $\bar\rho_{\mathbf{k}}|\psi_0\rangle$. Such SMA trial wavefunctions were shown to give an excellent description of the actual collective excitation for momenta smaller than the magnetoroton minimum~\cite{Repellin}, both for the exact Laughlin state and the Coulomb ground state. Generalizations of the SMA, based on Jack polynomials, have been shown to be microscopically accurate at \emph{all} values of momenta accessible in finite systems~\cite{yang2012model}. 

Since the SMA dominates the low-energy physics of an FQH state, it could be expected that it also plays a crucial role in the quench dynamics. However, this becomes less clear if one notices that our quench protocol preserves the translation symmetry of the initial state, hence all dynamics takes place in the $\mathbf{k}=0$ sector of the Hilbert space that contains the uniform Laughlin state. The $k\to 0$ limit of the SMA mode, i.e., the spin-$2$ graviton~\cite{yang2012model}, lies inside the 2 quasiparticle-2 quasihole continuum of states [Fig.~\ref{fig:quench}(b)], hence it is far from obvious that quench dynamics can be modelled by this single degree of freedom (although in some cases the $k\to 0$ limit of the SMA mode appears to be below the continuum of the energy spectrum \cite{jolicoeur2017shape}). In the following section, we present the results of our exact numerical simulations of the quench dynamics in finite-size systems. These results will show that the simplest type of geometric quench described by Eq.~(\ref{eq:ghat0}), indeed, gives rise to coherent oscillations of the geometric spin-2 degree of freedom for the Laughlin state. We note, however, that more general types of quenches are also possible, which cannot be mapped to a simple modification of the mass tensor. These quenches, which will be considered in Sec.~\ref{sec:higherspin}, excite  the tower of higher spin modes in the spectrum of an FQH state, giving rise to non-trivial dynamics, which is not captured by the simple graviton oscillation.

\section{Method}\label{sec:numerics}

\begin{figure}
\includegraphics[width=\linewidth]{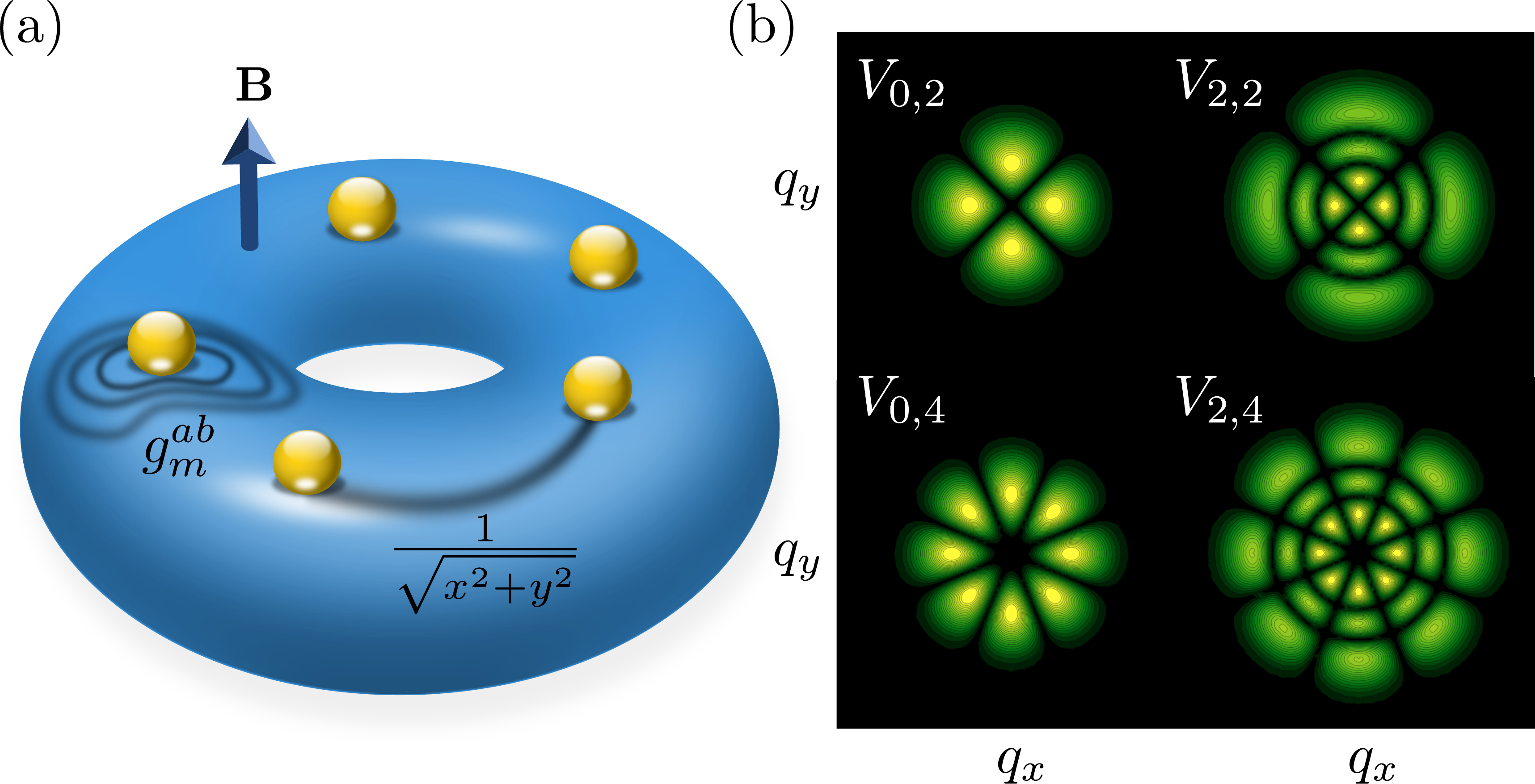}
\caption{(a) An FQH system on the surface of a torus threaded by a perpendicular magnetic field. The mass tensor $g_m$ determines the metric of the Landau level orbitals and is used to drive a quench, while the Coulomb potential is assumed to be isotropic.  (b) Contours in momentum space of the leading order anisotropic pseudopotentials for bosons~\cite{GeneralizedPPs} with quadrupolar ($V_{0,2}$, $V_{2,2}$) and octupolar ($V_{0,4}$, $V_{2,4}$) structure.}
\label{fig:torus}
\end{figure}

We model the quench by performing numerically exact time evolution with the Hamiltonian in Eq.~(\ref{eq:ham}) defined for $N$ particles and $N_\Phi$ flux quanta on a square torus [Fig.~\ref{fig:torus}(a)]. Magnetic translation symmetry~\cite{Haldane-PhysRevLett.55.2095} is used to reduce the complexity of the calculation and to classify the many-body eigenstates. We focus on the Laughlin states of bosons and fermions, corresponding to filling factors $\nu=1/2$ and $\nu=1/3$, respectively.

We consider two types of interaction potentials. In the first case, we assume a model (short-range) interaction for which the Laughlin state is the unique (and the densest) zero-energy state~\cite{Haldane1983}. This interaction is a simple contact repulsion for bosons, $V_\mathbf{q}=1$, and for fermions it is $V_\mathbf{q}=L_1(g_m^{ab}q_aq_b)$, where $L_1$ is the first Laguerre polynomial. As customary in the literature~\cite{HaldaneGeometry}, in the model interaction, we have assumed that the interaction metric $g_i$ coincides with the Landau orbit metric $g_m$. In the second case, we consider the Coulomb interaction, for which $g_i$ is kept isotropic, i.e., $V_\mathbf{q}=2\pi/\sqrt{q_x^2+q_y^2}$, but the mass tensor $g_m$ is allowed to be non-isotropic.

For both types of interaction, the quench is implemented by switching $g_m$ to $g_m'$ at time $t=0$, where both $g_m$ and $g_m'$, for simplicity, are taken to be diagonal, like in Eq.~(\ref{eg:iso}) below. For small system sizes, the subsequent time evolution is performed by obtaining all energy eigenvalues $\epsilon_n$ and the corresponding eigenvectors $|n\rangle$ of the quench Hamiltonian $H'$, and then evaluating
$ |\psi(t)\rangle = \sum_n \exp(-i \epsilon_n t)\langle n| \psi_0\rangle |n\rangle$. By restricting the sum to a subset of eigenstates, one can conveniently project the dynamics onto a desired energy shell. At larger system sizes, we use time-dependent Lanczos methods to iteratively compute $|\psi(t)\rangle$. 

We use the same parametrization for the intrinsic metric $\hat g$ as in Eq.~(\ref{eq:ghat0}). Due to its invariance under $Q\rightarrow -Q$ and $\phi\rightarrow \phi+\pi$, we focus on $Q\geq0$. For a given state $|\psi(t)\rangle$, $\hat g$ is determined by brute force search over a large set of (precomputed) trial states $|\psi_{\rm {trial}}\rangle$, i.e., model Laughlin states parametrized by $Q \in \left[0,Q_{\max}\right]$ and $\phi \in \left[ -\pi, \pi \right]$, such that the overlap $|\langle \psi_{\rm{trial}}|\psi(t)\rangle|$ is maximized. For adiabatic quenches, we found it sufficient to restrict to $Q_{\max}=1$ and $\phi\in[-\pi/2,\pi/2]$. The confidence in the result is quantified by the maximum overlap achieved: if this overlap is not close to 1 (like in the case of strong quench), none of the trial states appropriately describes the system and the intrinsic metric should be determined via some other means, for example, via minimization of the total energy or momentum polarization~\cite{GromovGeraedtsBradlyn}.

Finally, in order to identify states corresponding to the spin-$n$ excitation, we use  the recently developed formalism of anisotropic pseudopotentials~\cite{GeneralizedPPs}. In Ref.~\cite{GeneralizedPPs} it was shown that any two-particle interaction, including cases where metrics $g_m$ and $g_i$ are different, can be expanded into an orthonormal basis of generalized pseudopotentials. These operators maintain translation invariance, but break rotation symmetry to a discrete subgroup. Up to normalization prefactors, the generalized pseudopotentials are given by \cite{GeneralizedPPs}
\begin{eqnarray}
V_{m,n}\left(g_p; q_x, q_y \right) = L_m^n\left( g_p^{ab}q_a q_b\right) \textbf{q}^n + {\rm c.c.},\label{g1}
\end{eqnarray} 
where $m$, $n$ are even integers (for bosons). When $n=0$, these reduce to the usual (isotropic) Haldane pseudopotentials~\cite{Haldane1983}. $V_{m,n}$ explicitly depends on the metric $g_p$ through the argument of the Laguerre polynomial, as well as the vector $\mathbf{q}$. For simplicity, we fix $g_p=\id$ and $\mathbf{q}=\frac{1}{\sqrt{2}}(q_x+iq_y)$ in our numerics. 

Contours of leading-order $V_{m,n}$ for bosons are depicted in Fig.~\ref{fig:torus}(b). In particular, we can observe that the dominant pseudopotential $V_{0,2}$ has a clear quadrupolar structure in momentum space, indicating that $\hat{V}_{0,2}$ carries angular momentum $L_z=2$.
Based on this, we use $\hat {V}_{0,n}$ as a spectroscopic probe to identify many-body eigenstates, which are part of the effective spin-$n$ excitation. To that end, we introduce the following ``spectral function":
\begin{eqnarray}
\label{eq:spectralfunction}
I_{m,n}(\omega) = \sum_j \delta(\omega-\epsilon_j) |\langle j | \hat V_{m,n} |0\rangle|^2,
\end{eqnarray} 
where $\epsilon_j$ and $|j\rangle$ are the eigen-energy and eigenstate of Eq.~(\ref{eq:ham}).
This is a generalization of the spectral function that describes an acoustic wave absorption experiment~\cite{KunYangAcoustic}. The peaks of $I_{m,n}$ can be used to obtain the frequencies of different spin modes in the quench dynamics. In the following section, we investigate in detail $I_{0,2}$ and show that it indeed yields the frequency of the graviton mode.

\section{Results}\label{sec:results}

Here, we present our numerical results for adiabatic quenches, where the mass tensor is instantaneously changed from $g_m={\rm diag}(e^{A_0},e^{-A_0})$ to $g_m'={\rm diag}(e^{A_1},e^{-A_1})$ with $A_0$ and $A_1$ close to $0$. By ``adiabatic" quench we mean that the system remains in the Laughlin phase. Note that this does not imply that we restrict to an infinitesimal variation of  the parameters in the Hamiltonian. Indeed, the parameters of our quenched Hamiltonian differ by a finite amount from the original Hamiltonian, which allows the dynamics to explore a finite density of excitations.   

In the following, we focus on the dynamics of $\hat g$ for $\nu=1/2$ bosons in detail, while similar results for $\nu=1/3$ fermions are given in Appendix \ref{sec:app_fermions}. In all numerical data below, time is given in units of inverse energy [$e^2/(\epsilon\ell_B)$ in the case of Coulomb interaction], with $\hbar=1$.

\subsection{Isotropic-to-anisotropic quench}\label{sec:isotropicquench}

\begin{figure}[htb]
\includegraphics[width=\linewidth]{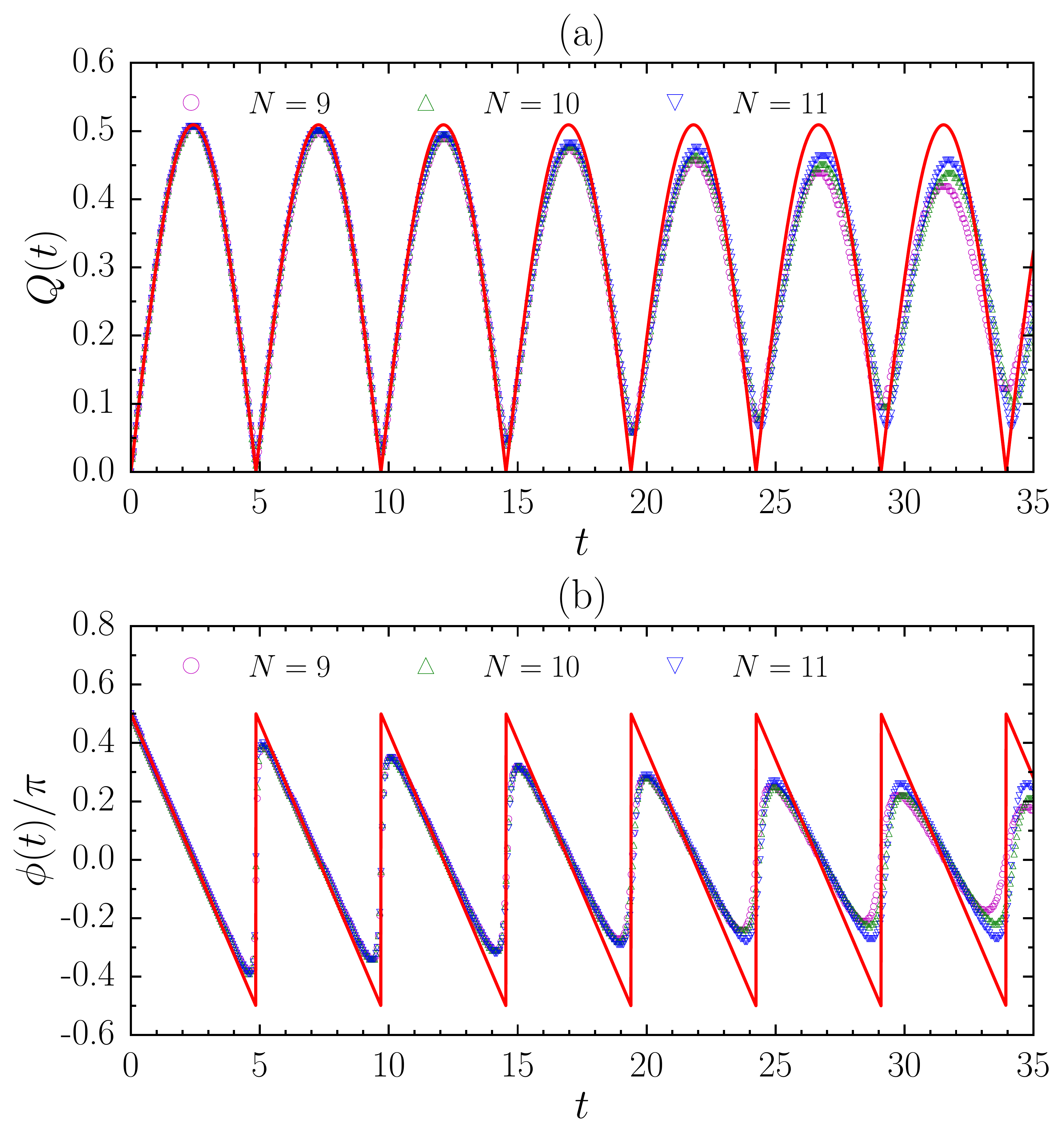}
\caption{Exact dynamics of (a) $Q$ and (b) $\phi$ for $N=9-11$ bosons at $\nu=1/2$ with the contact interaction. The quench is driven by choosing $A_0=0$ and $A_1=\ln 1.3$.  
The red curves are fits to Eqs.~(\ref{eq:linsol1}) and (\ref{eq:linsol2}), which coincide with the predictions of the bimetric theory, discussed in Sec.~\ref{sec:bimetric}.
}
\label{fig:v0_boson}
\end{figure}

\begin{figure}[htb]
\includegraphics[width=\linewidth]{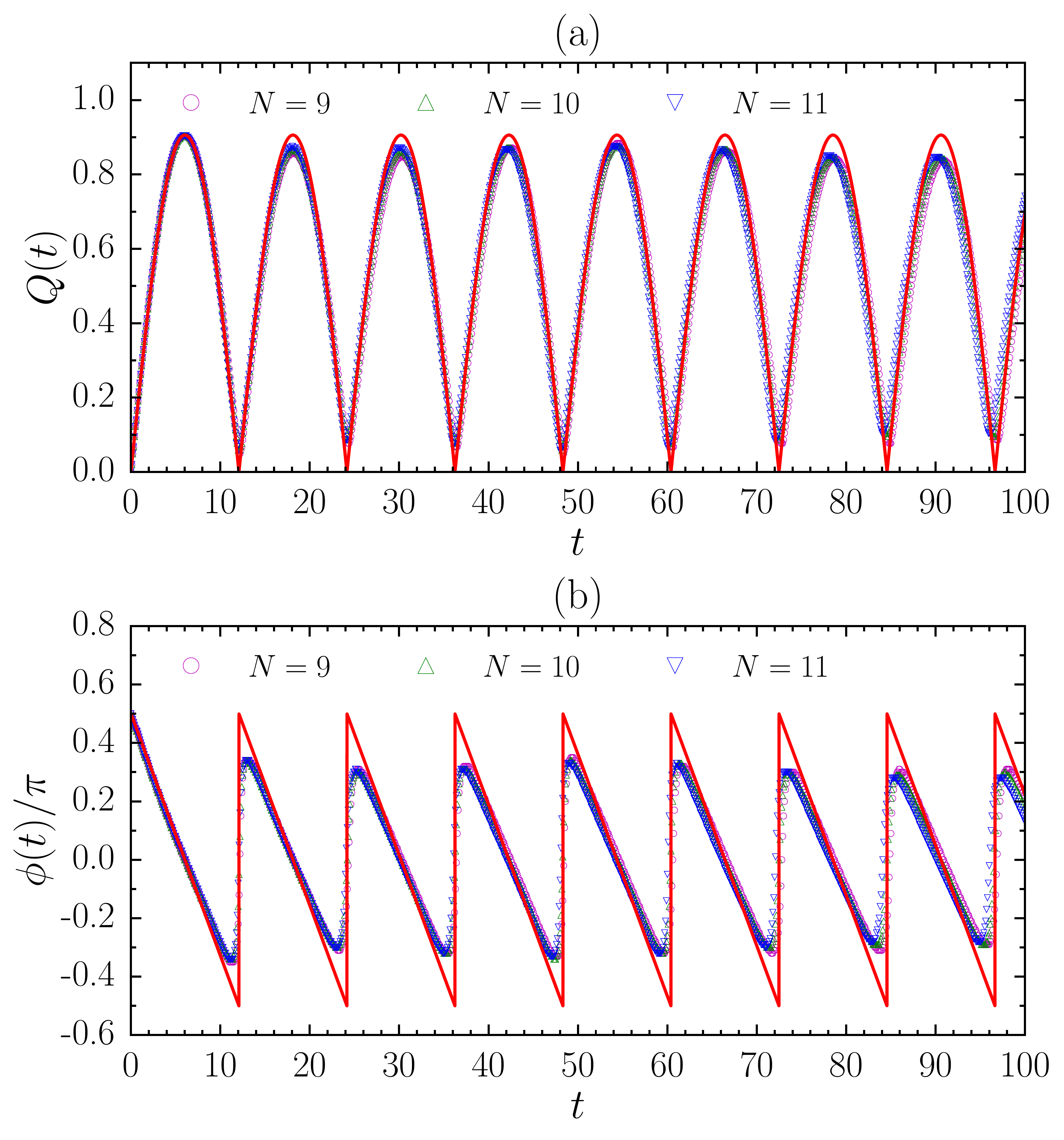}
\caption{Exact dynamics of (a) $Q$ and (b) $\phi$ for $N=9-11$ bosons at $\nu=1/2$ with the Coulomb interaction. The quench is driven by choosing $A_0=0$ and $A_1=\ln 2.0$.
The red curves are fits to Eqs.~(\ref{eq:linsol1}) and (\ref{eq:linsol2}), which coincide with the predictions of the bimetric theory, discussed in Sec.~\ref{sec:bimetric}.
}
\label{fig:coulomb_boson}
\end{figure}

We start with the simple case of $A_0=0, A_1>0$, where the initial state is the isotropic Laughlin state. 
As representative examples, we choose $A_1=\ln1.3$ and $A_1=\ln2.0$ for quenches driven by contact and Coulomb interactions, respectively. 
The intrinsic metric $\hat g$ of the post-quench state is determined by the maximum of the overlap between the post-quench state and a trial set of anisotropic Laughlin states.
The regime of adiabatic quench corresponds to the maximum of this overlap being close to unity.
The exact dynamics of $\hat g$ up to moderate times is shown in Figs.~\ref{fig:v0_boson} and \ref{fig:coulomb_boson} for the two types of interactions.
We observe that  both $Q$ and $\phi$ oscillate with a well-defined frequency (and in particular, $\phi$ appears to be a \emph{linear} function of time). This is rather surprising because the induced change in the microscopic structure of the FQH state is by no means small: for the above choices of $A_1$ in Figs.~\ref{fig:v0_boson} and \ref{fig:coulomb_boson}, we find the maximum anisotropy $e^Q$ of the post-quench state to be larger than the initial (isotropic) state by a factor $1.5-2.5$. 

Assuming a single harmonic, we might guess that the oscillations in Figs.~\ref{fig:v0_boson} and \ref{fig:coulomb_boson} are well described by
\begin{eqnarray}
Q(t) &=& 2A\sin\left(\frac{E_\gamma t}{2}\right),\;\;\; \phi(t) = \frac{\pi}{2} -\frac{E_\gamma t}{2},\label{eq:linsol1}\\
Q(t) &=& -2A\sin\left(\frac{E_\gamma t}{2}\right),\;\;\; \phi(t) = \frac{3\pi}{2}-\frac{E_\gamma t}{2},\label{eq:linsol2}
\end{eqnarray} 
where $Q$ obeys harmonic motion and $\phi$ has a simple linear dependence on time. Note that only one of these two solutions is independent because the system is invariant under $Q\rightarrow -Q$ and $\phi\rightarrow \phi+\pi$. Thus, we can focus on the $Q\geq0$ part and consider $\phi\;{\rm mod}\;2\pi$. By inspection of $Q(t)$, we see the solution will alternate between the two branches, which doubles the frequency from $E_\gamma/2$ to $E_\gamma$. The overall prefactor (written as $2A$) is expected to be proportional to the intrinsic anisotropy of the state. 

Fitting the first one or two oscillations in Figs.~\ref{fig:v0_boson} and \ref{fig:coulomb_boson} against Eq.~(\ref{eq:linsol1}) yields a remarkably accurate agreement with the full dynamics across the entire time interval. What sets the frequency, $E_\gamma$, of the oscillations? From the fits, we obtain $E_\gamma\approx 1.296$ and $0.520$ for the cases of contact and Coulomb interactions, respectively. These values do not match the spectral gap in the $\mathbf{k}=0$ momentum sector, which can be much smaller (see Fig.~\ref{fig:quench}). Instead, as shown below, they agree with the \emph{graviton} gap very well, which we \emph{independently} estimate to be $\approx 1.3$ for the contact interaction and $0.52$ for the Coulomb interaction. 

\begin{figure}[t]
\includegraphics[width=\linewidth]{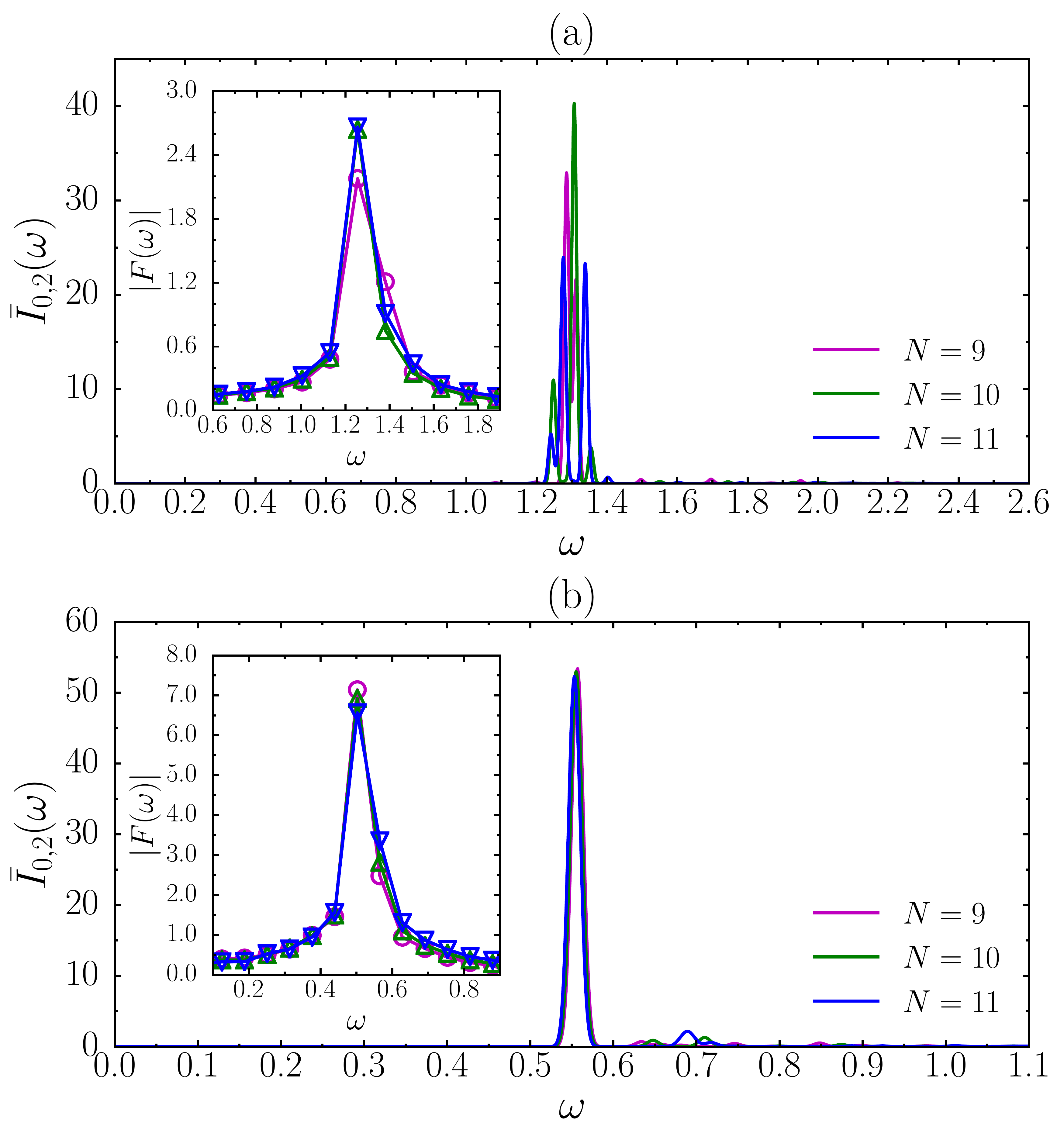}
\caption{Normalized spectral function ${\bar I}_{0,2}(\omega)=I_{0,2}(\omega)/\int I_{0,2}(\omega) d\omega$ for isotropic systems of $N=9-11$ bosons at $\nu=1/2$ with (a) contact and (b) Coulomb interactions. The insets show the discrete Fourier transform $|F(\omega)|$ of $Q$ in Figs.~\ref{fig:v0_boson} and \ref{fig:coulomb_boson}. 
Markers in insets and curves in main figures with the same color refer to the same system size. }
\label{fig:SF_FT}
\end{figure}

In order to justify the identification of $E_\gamma$ with the graviton gap, we use $I_{0,2}(\omega)$ spectral function defined in Eq.~(\ref{eq:spectralfunction}) to detect the states that carry spin-2. We expect peaks of $I_{0,2}(\omega)$ at energies corresponding to the spin-$2$ excitations that comprise the graviton mode. In Fig.~\ref{fig:SF_FT}, we show $I_{0,2}(\omega)$ [normalized by $\int I_{0,2}(\omega) d\omega$] for contact and Coulomb interactions in isotropic systems (similar data are obtained for weakly anisotropic systems). In the case of contact interaction, we observe sharply pronounced peaks of $I_{0,2}(\omega)$ around $\omega=1.3$ for various system sizes [Fig.~\ref{fig:SF_FT}(a)], which means that spin-$2$ eigenstates concentrate in this narrow energy window, corresponding to the graviton gap $E_\gamma \approx 1.3$. For Coulomb interacting systems, a pronounced peak in $I_{0,2}(\omega)$ also exists, but at a lower energy $\omega\approx 0.55$ [Fig.~\ref{fig:SF_FT}(b)].  As an independent estimate of the graviton gap, we use the discrete Fourier transform of $Q$ (see insets of Fig.~\ref{fig:SF_FT}). This Fourier transform also shows a sharp peak around the same energy as that in $I_{0,2}(\omega)$, which further confirms that the dynamics of the intrinsic metric is indeed caused by the graviton oscillation.

\subsection{States contributing to dynamics}

The agreement between the oscillation frequency in the exact dynamics of $\hat g$ and the graviton gap strongly implies that the dynamics is dominated by spin-$2$ eigenstates of $H'$. We now scrutinize this conjecture by quantifying how well the full dynamics can be reproduced by an explicit projection to the subset of eigenstates centered around the energy of the spin-$2$ mode.

\begin{figure}[htb]
\includegraphics[width=\linewidth]{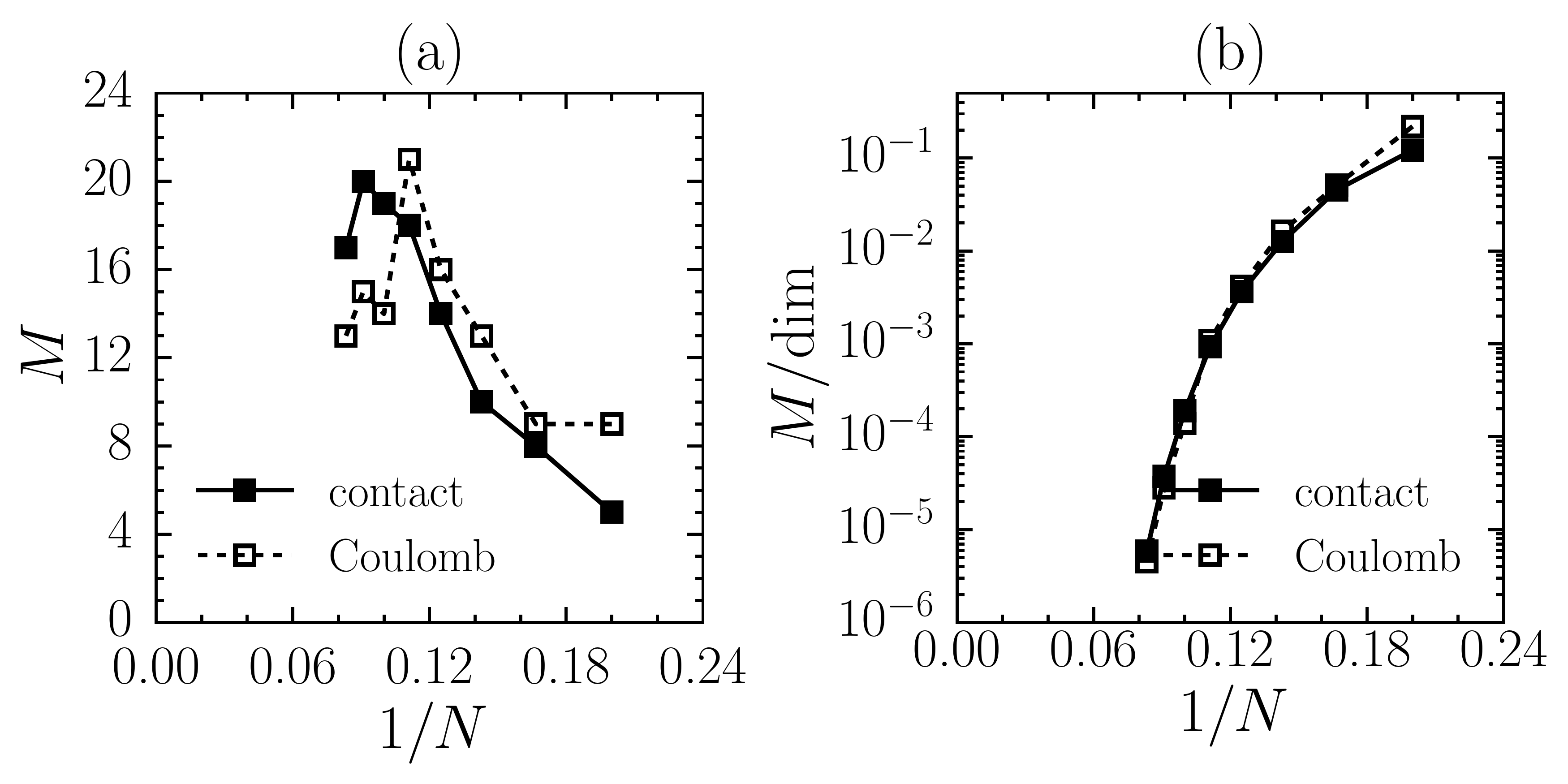}
\caption{Finite-size scaling of the number of spin-$2$ states in the ${\bf k}=0$ Hilbert space for $N=5-12$ bosons at $\nu=1/2$. $M$ represents the number of states which contain $f=99\%$ of the total $|\langle n| \hat V_{0,2} |0\rangle|^2$. We consider both contact and Coulomb quench Hamiltonians, with $A_1=\ln 1.3$ and $A_1=\ln2.0$, respectively. The plots show (a) $M$ and (b) $M/{\rm dim}$, where ${\rm dim}$ is the ${\bf k}=0$ Hilbert space dimension.
}
\label{fig:v02_weight}
\end{figure}

The projected subspace is defined by the eigenstates $|n\rangle$ of $H'$ with dominant matrix elements $|\langle n| \hat V_{0,2} |0\rangle|^2$. The weight of this projection is measured by 
\begin{eqnarray}
f=\left(\sum_{n=0}^{M-1} |\langle n| \hat V_{0,2} |0\rangle|^2\right)/\left(\sum_{n=0}^{D-1} |\langle n| \hat V_{0,2} |0\rangle|^2\right),
\end{eqnarray}
where $|n\rangle$ have been sorted in descending order according to $|\langle n| \hat V_{0,2} |0\rangle|^2$. Here $M$ is the number of kept states and $D$ is the total number of eigenstates that we can obtain from exact diagonalization of $H'$, i.e., $D$ is equal to the Hilbert space dimension in the ${\bf k}=0$ sector for $N\leq 9$, while for practical reasons we kept $D=200$ for larger systems ($N=10-12$). Remarkably, the number of states we need to saturate the total weight $f$ is significantly smaller than the Hilbert space dimension. Fixing $f=99\%$, the finite-size scaling of the required $M$ clearly indicates that only an exponentially small fraction of the whole Hilbert space carries spin-$2$, as seen in Fig.~\ref{fig:v02_weight}. 

While the exponentially decaying ratio of $M$ to the Hilbert space dimension might not be entirely surprising (since it still does not preclude $M$ being exponentially large in the system size), it can also be shown that, in absolute terms, only a very small number of states $M$ yields a very accurate approximation to the exact dynamics. In Appendix~\ref{sec:projected}, we perform the above projection, based on the $|\langle n| \hat V_{0,2} |0\rangle|^2$ matrix elements, for the dynamics of $N=11$ bosons driven by the isotropic-to-anisotropic quench (Fig.~\ref{fig:project_boson}). Remarkably, the main features of the exact dynamics are already accurately reproduced for $M$ less than $10$, even if there are as many as $533160$ eigenstates in the entire ${\bf k}=0$ Hilbert space. Such an excellent approximation of the exact dynamics by an exponentially small fraction of states in the Hilbert space is further evidence for the picture of graviton oscillation.

\section{Adiabatic geometric quench in bimetric theory}\label{sec:bimetric}

In the previous Section, we have established that quenching the mass tensor, Eq.~(\ref{eq:ghat0}), in the Laughlin phase gives rise to dynamics which is supported over a vanishingly small fraction of states that carry spin-2. Thus we expect that our results in Sec.~\ref{sec:results} can be used to test the validity of the bimetric theory~\cite{GromovSon}. Here, we generalize the bimetric theory to anisotropic quantum Hall states, and derive an analytical description of the dynamics of the intrinsic metric $\hat g$ after a quench exciting spin-$2$ states. We find that this analytical prediction agrees remarkably well with the exact numerics of geometric quench in Sec.~\ref{sec:results}, which supports the validity of the bimetric theory. Despite of its agreement with the bimetric theory in special quenches exciting spin-$2$ states, we should emphasize that geometric quench is a logically independent method and contains much new physics beyond the bimetric theory (see Sec.~\ref{sec:higherspin}).

\subsection{Lagrangian for the isotropic case}

The bimetric theory describes the gapped dynamics of a single spin-$2$ degree of freedom in an FQH system, interacting with an external electro-magnetic field and ambient geometry. The dynamical degree of freedom is the vielbein $\hat e^\alpha_i$ \cite{carroll2004spacetime} that ``squares'' to the dynamic, unimodular metric $\hat g_{ij} = \hat e^\alpha_i\hat e^\beta_j \delta_{\alpha \beta}$, or simply $\hat{g} = \hat{e} \cdot \hat{e}^T$. Here, $\hat g$ corresponds to the intrinsic metric of the FQH state mentioned in Sec.~\ref{sec:quench}. The inverse metric is given by $\hat G^{ij} = \hat E_\alpha^i\hat E_\beta^j \delta^{\alpha \beta}$, where $\hat E_\alpha^i$ is the inverse vielbein \cite{carroll2004spacetime}, satisfying $\hat E^i_\alpha \hat e_j^\alpha = \delta^i_j$. The unimodular condition on $\hat g_{ij}$ takes form $\sqrt{g} = \sqrt{\hat g}$, where $\sqrt{g}$ is the determinant of the \emph{ambient} metric $g_{ij} = \delta_{AB} e^A_i e^B_j$ and $e^A_i$ is the ambient vielbein. Thus, in flat ambient space (that is, when $g_{ij} = \delta_{ij}$), $\sqrt{\hat g} = 1$.

In the absence of an external electric field, and when the magnetic field is homogeneous, the Lagrangian takes the form \cite{GromovSon}
\be\la{eq:isoL}
\mathcal L  = \frac{\nu \varsigma}{2\pi \ell_B^2}  \hat \omega_0 - \frac{m}{2} \left[ \frac{1}{2}\hat g_{ij} g^{ij} - \gamma \right]^2 = \mathcal L_{\rm top} + \mathcal L_{\rm pot}\,,
\ee 
where $\hat \omega_0$ is the temporal component of the (dynamic) Levi-Civita spin connection, given by
\be
\hat \omega_0 = \frac{1}{2} \epsilon_{\alpha}{}^\beta \hat E^i_\beta \partial_0  \hat e_i^\alpha\,.
\ee
The phenomenological coefficient $m$ sets the energy scale that determines the gap of the spin-$2$ mode.
The quantized coefficient $\varsigma$ is determined by the ``shift" $\mathcal S$ \cite{GromovSon} and takes the value $|\varsigma| = \frac{|\mathcal S - 1|}{2}$. The phenomenological parameter $\gamma$ is used to tune the theory close to the nematic phase transition, where the SMA is exact, in the gapped phase $\gamma < 1$. Further details on the bimetric Rieman-Cartan geometry can be found in Refs.\,\cite{GromovSon, GromovGeraedtsBradlyn}. We will utilize it to introduce anisotropy into the problem.

\subsection{Anisotropy in bimetric theory}
To introduce the quadrupolar anisotropy from a non-trivial mass tensor or other sources, we take the inspiration from Ref.\,\cite{GromovGeraedtsBradlyn}, where quadrupolar anisotropy was described in geometric terms. More concretely, we introduce a unimodular matrix $\mathrm{m}_{AB}$, and construct a rank-2 tensor from it according to
\be
g^{(\mathrm{m})}_{ij} = \mathrm{m}_{AB} e^A_i e^B_j\,,
\ee
where $e^A_i$ are the vielbeins that describe the ambient geometry. We now generalize Eq.~\eqref{eq:isoL} to the anisotropic case. To do so, we simply replace $g_{ij}$ by $g^{(\mathrm{m})}_{ij}$ in Eq.~\eqref{eq:isoL}.

To analyze the theory further, we make the following simplifying assumptions: (i)  the ambient space is assumed to be flat, $e^A_i = \delta^A_i$, and (ii) we assume a particular parametrization of $g^{(\mathrm{m})}_{ij}$, given by
\begin{eqnarray}\label{eg:iso}
g_{ij}^{(\mathrm{m})} =\left(
 \begin{array}{cc}
 e^A & 0 \\
 0 & e^{-A} \\
\end{array}
\right)\,,
\end{eqnarray}
where $A$ is a real parameter that defines the effective anisotropy in the FQH system, which is a compromise between two extrinsic metrics $g_m$ and $g_i$ as discussed in  Sec.~\ref{sec:quench}. The abrupt change of the mass tensor $g_m$ in our quench protocol will lead to the change of $A$ in Eq.~(\ref{eg:iso}). 

In order to compute the dynamics of $\hat g_{ij}$, we parametrize it in terms of $Q$ and $\phi$ as in Eq.~(\ref{eq:ghat0}).
Both $Q$ and $\phi$ are functions of time, but not space, since the problem we will consider is homogeneous (i.e., the quench is global). The two terms in the Lagrangian take the form
\begin{eqnarray}
\mathcal L_{\rm top} &=& \frac{\varsigma \bar \rho}{2} \left( 1 - \cosh Q \right) \dot \phi\,,
\\
\mathcal L_{\rm pot} &=& -\frac{m}{2}  \left( \gamma+\sinh A \sinh Q \cos \phi -\cosh A \cosh Q \right)^2\;.\;\;\;\;\;\;
\end{eqnarray}
The equations of motion are found from
\be
\frac{\delta S}{\delta \phi} = 0 \,, \qquad \frac{\delta S}{\delta Q} = 0\,,
\ee
where $S = \int d^3x \,\mathcal L$ is the action. 
The equation for $\phi (t)$ takes the form
\begin{widetext}
\bea  \la{eq:phi}
 \dot{\phi} \sinh Q = - 2 \Omega \left( \sinh A \cosh Q \cos \phi -\cosh A \sinh Q\right)  \left(\gamma+\sinh A \sinh Q \cos \phi -\cosh A \cosh Q \right)\,,
\eea
and for $Q(t)$ we find
\bea \la{eq:Q}
\dot Q \sinh Q = - 2 \Omega \sin \phi \sinh Q \sinh A   \left( \gamma+\sinh A \sinh Q \cos \phi -\cosh A \cosh Q \right)\,,
\eea
where we have introduced the energy scale $\Omega =  \frac{m}{ \bar \rho \varsigma}$. Equations\,\eqref{eq:phi} and \eqref{eq:Q} are the central result of this section. 

\end{widetext}

\subsection{Isotropic limit}

In the isotropic case we must take $A \rightarrow 0$. We find
\be
\dot Q = 0\,, \qquad \dot \phi = 2\Omega \left( \gamma - \cosh Q \right)  \, .
\ee
Choosing the solution $Q=0$ of the first equation, the second equation reads 
\be
\dot \phi = - 2 \Omega ( 1 - \gamma) \equiv -E_\gamma\, \quad  \Rightarrow \quad \phi(t) = \phi(0)- E_\gamma t\,,
\ee
from where we interpret $E_\gamma = 2\Omega(1-\gamma) = \frac{2 m (1-\gamma)}{\bar \rho \varsigma}$ as the gap of the spin-$2$ part of the GMP mode at $\mathbf k =0$~\cite{GromovSon}. When $\gamma \rightarrow 1$, the gap closes and the FQH state undergoes a nematic phase transition~\cite{GromovSon}. The dynamical metric evaluated on this solution takes the form
\be
\hat g_{ij} = \left(
\begin{array}{cc}
 1& 0 \\
 0 & 1 \\
\end{array}
\right)\,,
\ee
consequently the dynamics of $\phi(t)$ is not visible in the fluctuations of the dynamical metric. Formally speaking, this happens because $Q=0$. When $Q$ is different from $0$ the metric will be sensitive to the dynamics of $\phi$. This is what happens in the case of weak anisotropy discussed next.

\subsection{Adiabatic geometric quench in bimetric theory}

In order to perform the quench, we have to phrase it in terms of Eqs.~\eqref{eq:phi} and \eqref{eq:Q}. Taking inspiration from Ref.~\cite{franchini2015universal}, we perform the quench in two steps. First, we need to fix the initial condition $Q(0),\phi(0)$ for Eqs.~\eqref{eq:phi} and \eqref{eq:Q}, which should be determined by the intrinsic metric of the initial Laughlin state $|\psi_0\rangle$. Second, to obtain the quench dynamics, we solve Eqs.~\eqref{eq:phi} and \eqref{eq:Q} under such initial condition, with the value of $A$ determined by the effective anisotropy of $H'$ (or equivalently, by the intrinsic metric of the ground state of $H'$). 

When anisotropy is weak, we can assume both $A$ and $Q$ are close to $0$. Taylor expansion of Eqs.~\eqref{eq:phi} and \eqref{eq:Q} in $A$ and $Q$ leads to the following system of linear equations
\bea\la{eq:phidotweak}
&&\dot \phi Q= E_\gamma\left( A \cos \phi-Q \right) \,, 
\\ \la{eq:Qdotweak}
&& \dot Q =  E_\gamma A \sin \phi\,, 
\eea
for which analytical solutions exist under suitable initial conditions. In particular, for $Q(0)=0$, it can be verified by direct substitution that our earlier Eqs.~(\ref{eq:linsol1})-(\ref{eq:linsol2}), which were used to fit the numerical data in Sec.~\ref{sec:results}, are exact solutions.

\begin{figure}[htb]
\includegraphics[width=\linewidth]{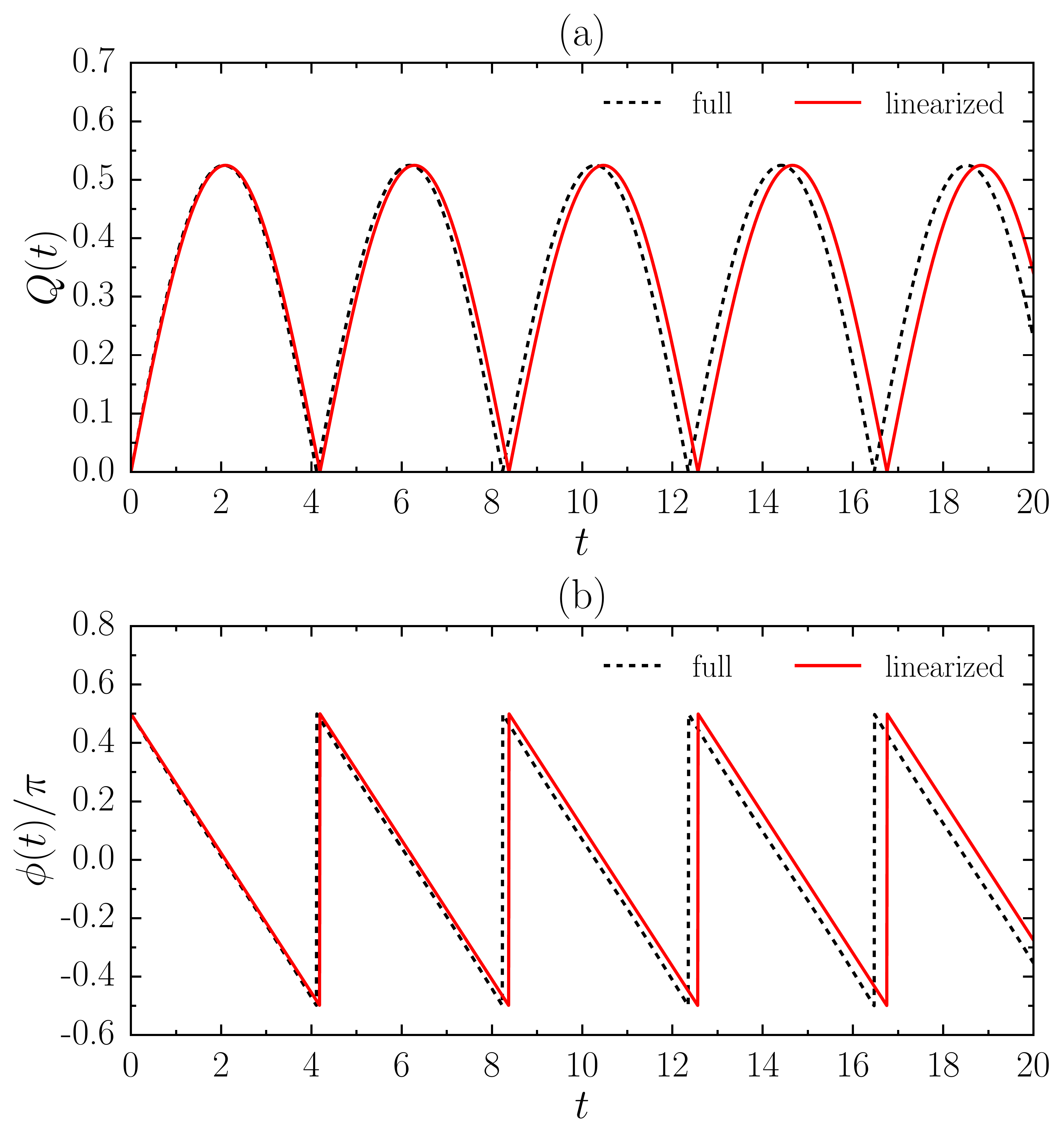}
\caption{Predictions of the bimetric theory for (a) $Q$ and (b) $\phi$ under the initial condition $Q(0)=0$. Black dashed lines indicate the numerical solution of the full equations, Eqs.~(\ref{eq:phi}) and (\ref{eq:Q}). Red solid lines indicate the analytical solution (\ref{eq:linsol1}) and (\ref{eq:linsol2}) of the linearized system, Eqs.~\eqref{eq:phidotweak} and \eqref{eq:Qdotweak}. Here, we focus on the $Q\geq0$ solution.
The parameters are fixed at $\Omega=1.5/4$, $\gamma=-1$, $A=\ln1.3$.}
\label{fig:solution_linear}
\end{figure}

As long as the quench is adiabatic, the linearized Eqs.~\eqref{eq:phidotweak} and \eqref{eq:Qdotweak} provide a very accurate approximation to the full non-linear dynamics. Fig.~\ref{fig:solution_linear} shows the comparison of the numerical solutions of non-linear Eqs.~\eqref{eq:phi} and \eqref{eq:Q} satisfying $Q(0)=0$ against their linearized counterparts, Eqs.~(\ref{eq:linsol1}) and (\ref{eq:linsol2}), where we focus on the $Q\geq0$ part and consider $\phi\;{\rm mod}\;2\pi$. Red solid lines in Fig.~\ref{fig:solution_linear} result from a combination of Eqs.~(\ref{eq:linsol1}) and (\ref{eq:linsol2}), which gives us $Q$ and $\phi$ as periodic functions with frequency $E_\gamma$.  Mathematically, it is rather surprising that Eqs.~(\ref{eq:linsol1})-(\ref{eq:linsol2}) display such close agreement with the solutions of Eqs.~\eqref{eq:phi} and \eqref{eq:Q}, since the latter appear to be strongly non-linear.

\subsection{Agreement between bimetric theory and exact dynamics}\label{sec:bimetric_numerics}

Since Eqs.~\eqref{eq:phidotweak}-\eqref{eq:Qdotweak} are an accurate approximation of Eqs.~\eqref{eq:phi}-\eqref{eq:Q} in the adiabatic quench regime, we were justified in fitting our exact results in Sec.~\ref{sec:results} against simple harmonic oscillation in Eqs.~(\ref{eq:linsol1})-(\ref{eq:linsol2}).  The fits allow us to extract the values of $A$ and $E_\gamma$, which we can now interpret as the intrinsic anisotropy and graviton gap in the bimetric theory. 

By extracting $E_\gamma$ from the fitting of quench data for various initial parameters, we indeed find that the obtained $E_\gamma$ is  insensitive to the precise values of $A_0$ and $A_1$, as long as we are in the adiabatic quench regime. On the other hand, $A$  quantifies the intrinsic anisotropy in the ground state of $H'$, which is a compromise between $g_m'$ and $g_i'$. Since we do the same modification for $g_m$ and $g_i$ in the quench driven by the contact interaction, we expect $A=A_1$ in this case. Indeed, we get $A\approx 0.255$ from the fit in Fig.~\ref{fig:v0_boson}, which is very close to $A_1=\ln 1.3\approx 0.262$. However, in the quench driven by the Coulomb interaction, we keep isotropic $g_i$ and only change $g_m$, so the intrinsic anisotropy in the ground state of $H'$ is expected to be weaker than $g_m'$. The fit in Fig.~\ref{fig:coulomb_boson} gives $A\approx 0.454$, which is indeed between $0$ and $A_1=\ln2.0\approx 0.693$. 

In summary, we find that adiabatic isotropic-anisotropic mass anisotropy quenches in the Laughlin phase can be described as harmonic motion of a single spin-$2$ degree of freedom, in agreement with the bimetric theory. Minor deviations from harmonic motion can be observed in Figs.~\ref{fig:v0_boson} and \ref{fig:coulomb_boson} at longer times. These deviations generally have two manifestations: as a decay in amplitude while maintaining the overall oscillation structure, or as a total departure from the oscillation. The former is mainly the case for isotropic-to-anisotropic quenches considered in Figs.~\ref{fig:v0_boson} and \ref{fig:coulomb_boson}, while the latter can be observed in the more general case of anisotropic-to-anisotropic quench considered in Appendix~\ref{sec:anisotropic} as well as in some fermionic data in Appendix~\ref{sec:app_fermions}. 

There are several possible explanations for the discrepancy with the bimetric theory. First, in finite systems a ``fragmentation" of the spin-$2$ graviton mode into several states with large matrix element $|\langle n | \hat V_{0,2} |0\rangle|^2$ (as seen in Fig.~\ref{fig:SF_FT}) induces several close frequencies in the dynamics. With the available data for system sizes up to $N=12$, there is no clear indication that this fragmentation disappears in the thermodynamic limit, even though a ``sharp" spin-$2$ mode is theoretically anticipated in large systems. This suggests that finite size effects, to the zeroth order, do not seem to be the explanation
for the fragmentation. Note that the fragmentation for the Coulomb interaction is much weaker than that for the contact interaction (Fig.~\ref{fig:SF_FT}), reflecting a more isolated graviton mode in the Coulomb case. A more likely source of ``finite-time" errors in the dynamics could be due to an effective Lieb-Robinson~\cite{LiebRobinson} ``light cone" exceeding the finite size of the system. 
Finally, the discrepancy could arise due to the contribution of higher-spin modes to the dynamics. In the following section, we design more complex type of quenches to probe such modes. These quenches lead to non-trivial dynamics which is not accounted for in the bimetric theory or its direct extensions in Eqs.~(\ref{eq:linsol1})-(\ref{eq:linsol2}).

\begin{figure*}[htb]
\includegraphics[width=\linewidth]{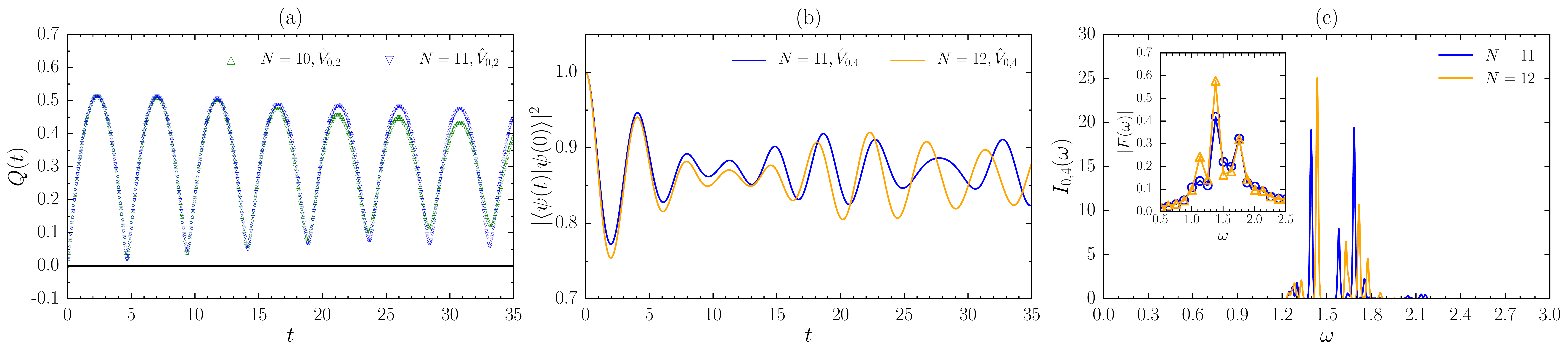}
\caption{(a) Exact dynamics of $Q$ for the quench with $\hat{V}_0\rightarrow\hat{V}_0+0.1\hat{V}_{0,2}$ (color symbol) and $\hat{V}_0\rightarrow\hat{V}_0+0.1\hat{V}_{0,4}$ (black solid line). A complete absence of dynamics of $\hat g$ can be noticed in the case of the perturbation by $\hat{V}_{0,4}$. 
(b) Dynamics of quantum fidelity, $|\langle\psi(t)|\psi(0)\rangle|^2$, for the quench with $\hat{V}_0\rightarrow\hat{V}_0+0.1\hat{V}_{0,4}$.
(c) Normalized spectral function ${\bar I}_{0,4}(\omega)=I_{0,4}(\omega)/\int I_{0,4}(\omega) d\omega$ for isotropic systems of $N=11$ and $12$ bosons at $\nu=1/2$ with contact interaction. The inset shows the discrete Fourier transform $|F(\omega)|$ of $|\langle\psi(t)|\psi(0)\rangle|^2$. 
Markers in insets and curves in main figures with the same color refer to the same system size.
}
\label{fig:v0204_boson}
\end{figure*}

\section{Higher-spin modes}\label{sec:higherspin}

\subsection{$W_\infty$ algebra in the lowest Landau level}

It has been realized a long time ago that the lowest Landau level admits an action of the (infinite-dimentional) algebra of area-preserving diffeomorphisms (APD), $W_\infty$ \cite{GMP86, cappelli1993infinite}. The action of this algebra can be interpreted as area-preserving distortions of FQH droplets, illustrated in Fig.~\ref{fig:quench}. These distortions, however, cost finite energy since the interaction potential, Eq.~\eqref{eq:ham}, is not invariant under these transformations~\cite{GMP86, HaldaneGeometry}. Consequently, the gapped spin-$2$ mode discussed in the previous section can be viewed as a particular, quadrupolar, APD of the droplet. This fact is reflected in the bimetric theory, in the form of hidden $sl(2,\mathbb R)$ (which is the largest finite-dimensional subalgebra of $W_\infty$) symmetry~\cite{GromovSon}. The topological part of the action in Eq.~\eqref{eq:isoL} is invariant under the action of $sl(2,\mathbb R)$, while the Hamiltonian explicitly breaks it down to $so(2)\subset sl(2,\mathbb R)$. Thus the application of the (broken) shear generators of $sl(2,\mathbb R)$, represented by the off-diagonal components of the emergent metric $\hat g_{ij}$, leads to the creation of an excited spin-$2$ state. 

Naturally, one may expect that higher angular momentum distortion of the fluid will correspond to independent collective modes, at least at long wavelengths~\cite{Cappelli2016, golkar2016higher}. The construction of an effective theory describing the infinite number of interacting higher spin fields is not an easy task, and has been an outstanding problem for a long time. A significant advance in solving this problem in $(2+1)D$ was made in~\cite{blencowe1989consistent, bergshoeff1990area}, where Chern-Simons-type action for higher spin gravity was constructed. Presently, it is not understood how these ideas can be applied to the FQH problem.

The algebra of APDs made multiple appearances in the FQH problem: non-commutative Chern-Simons theory \cite{susskind2001quantum}, matrix models \cite{polychronakos2001quantum, hellerman2001quantum}, bosonization of the LLL problem~\cite{iso1992fermions}, manifestly LLL-projected description of the CFL-type state of bosons at $\nu=1$~\cite{read1998lowest}.  Higher-spin degrees of freedom naturally appear in the composite fermion approach to the Jain series at filling $\nu=\frac{n}{2n+1}$, at large values of $n$ (i.e., close to half-filling)~\cite{golkar2016higher, nguyen2017fractional}. Those theories describe free massive higher spin fields with a quadratic action. Consequently, the (inherently non-linear) $W_\infty$ structure is invisible. In what follows we provide evidence, obtained directly from the microscopic quench dynamics, that at long distances the electrons forming an FQH fluid support collective modes of higher spin. The evidence is obtained by designing a quench protocol that selectively excites the higher-spin  collective modes.

\subsection{Higher-spin quench}

Tuning the anisotropy of the mass tensor is an experimentally relevant mechanism for changing the metric in the interaction. However, this also leads to the modification of the entire set of coefficients $c_{m,n}$ of the generalized Haldane pseudopotentials $\hat{V}_{m,n}$ [Eq.~(\ref{g1})], which uniquely characterize the interaction: 
\begin{eqnarray}
\bar V_{\mathbf{q}} = \sum_{m,n}c_{m,n}V_{m,n}(\mathbf{q}).
\end{eqnarray}
Modifying individual coefficients $c_{m,n}$, one at a time, is a much more controlled way to drive a quench. While this may not be easily achievable in experiment, this new quench protocol allows for more control  in probing different types of dynamics. As we show below, reformulating the quench protocol in this way not only includes the former scenario of changing the mass tensor, but also reveals new types of dynamics, which lie beyond the single mode approximation and the bimetric theory.

To perform the quench, we prepare a state annihilated by the Hamiltonian with $c_{m,n} =\delta_{n,0}\delta_{m,0}$. Then, a nonzero $c_{0,2}=0.1$ is introduced at $t=0^+$. This corresponds to the modification of the interaction $\hat{V}_0 \to \hat{V}_0+0.1\hat{V}_{0,2}$. Since $\hat{V}_{0,2}$ carries angular momentum $L_z=2$, we expect that its presence will drive a quench dominated by the spin-$2$ mode. Indeed, we find that the consequent exact dynamics of $\hat g$  behaves according to the description based on the spin-$2$ graviton oscillation [Fig.~\ref{fig:v0204_boson}(a)], and is very similar to that obtained by changing the mass tensor in Fig.~\ref{fig:v0_boson}.

Next, we consider a quench driven by adding a small amount of $\hat{V}_{0,4}$ component ($c_{0,4}=0.1$) such that the interaction changes according to $\hat{V}_0 \to \hat{V}_0+0.1\hat{V}_{0,4}$ at $t=0^+$. This type of quench can be realized in systems whose Fermi contours are not simple ellipses, e.g., where the band dispersion has the form $\epsilon(\mathbf{k}) = k^N \cos(N\theta)$ \cite{HaldaneShen, IppolitiNfold}. As shown in Ref.~\cite{IppolitiNfold}, when $N=4$, the dominant anisotropic pseudopotential is $V_{0,4}$. 
Surprisingly, we find $Q(t)\equiv0$ during the entire measured time interval [see the black solid line in Fig.~\ref{fig:v0204_boson}(a)], which suggests that the spin-$2$ collective mode is not excited. However, the system \emph{does} respond to the perturbation by $\hat V_{0,4}$, and we conjecture that this happens by exciting the higher-spin collective modes, which cannot be described by the bimetric theory. While we cannot describe the dynamics of higher-spin modes analytically, we can measure the dynamics of quantum fidelity, $|\langle \psi(t) | \psi(0) \rangle|^2$, shown in Fig.~\ref{fig:v0204_boson}(b). Clear oscillations in the fidelity are a signature of the non-trivial microscopic dynamics taking place.  We note that in the case of $\hat V_{0,2}$ quench, fidelity also oscillates, with the same frequency as $\hat g$. Moreover, we emphasize that our choice of measuring fidelity was merely a convenience; indeed, observables like entanglement entropy also display quantitatively similar oscillations.

The failure of the bimetric theory to capture the quench dynamics driven by $\hat{V}_{0,4}$ strongly suggests that the $\hat V_{0,4}$ quench is dominated by a higher-spin collective mode rather than the spin-$2$ one. Indeed, unlike the quadrupolar $\hat{V}_{0,2}$, $\hat{V}_{0,4}$ has octupolar structure in momentum space [Fig.~\ref{fig:torus}(b)], which implies that it would couple the ground state to higher-spin excitations. In our case, this is a spin-4 excitation because the odd-spin ones vanish due to inversion symmetry, which is a generic symmetry of LLL-projected FQH Hamiltonians. The spin-4 character of the mode is explicitly confirmed by evaluating the spectral function $\bar I_{0,4}(\omega)$ in Fig.~\ref{fig:v0204_boson}(c) (with identical results obtained for the discrete Fourier transform, shown in the inset of the same figure). The spectral function shows two peaks, one of which corresponds to the graviton energy and an additional one at higher energy, $\omega \sim 1.7$, which we identify with spin-4 mode. This provides an unambiguous example in which the $2\times 2$ unimodular metric in Eq.~(\ref{eq:ghat0}) is inadequate to capture the dynamics of the Laughlin state, motivating the generalization of the geometrical description of the fractional quantum Hall states and the effective field theory.

\section{Conclusions and outlook}\label{sec:conclusions}


In the present manuscript we have laid out the theoretical foundation for studying the nonequilibrium dynamics of FQH states. We have proposed and numerically simulated the geometric quench protocol that excites neutral collective degrees of freedom of  FQH liquids, such as  the Laughlin states of bosons at $\nu = 1/2$ and of fermions at $\nu = 1/3$. In all the cases, for the simplest types of quenches, which are driven by modifying the mass tensor, we have established that the short-time dynamics after the quench is dominated by the spin-$2$ graviton mode and accurately agrees with the bimetric theory. Furthermore, we have demonstrated that the geometric quench protocol admits a generalization that allows one to excite higher-spin collective degrees of freedom, whose dynamics cannot be described by the  currently available geometric theories of the FQH effect. We believe that the present work will motivate a more detailed study of the geometric degrees of freedom in various strongly correlated systems, and below we highlight several directions that open up for future investigations.

The immediate question that arises is the experimental applicability of our results. Even though the quench protocol can in principle be directly implemented with the available technology, the main challenge is observing coherent dynamics in solid-state materials that currently support FQH states. Assuming typical magnetic fields $10$T, we estimate the required time scales for observing the graviton oscillation to be $\sim 10^{-14}$s, which is prohibitively small. The second challenge is how to measure the intrinsic metric in \emph{gapped} FQH states. Recent experiments have successfully measured the anisotropy of the composite fermion Fermi surface induced by tilting the magnetic field~\cite{Kamburov,Kamburov2}. Similar measurements were also performed for hole systems~\cite{Mueed}. However, such experiments still need to be generalized to gapped FQH states (we note that acoustic wave absorption has been theoretically proposed as an alternative spectroscopic probe of the graviton~\cite{KunYangAcoustic}). Given the smallness of time scales in solid state FQH materials, a more versatile platform to observe geometric quench could be their analogs in lattice systems -- the fractional Chern insulators (FCIs) \cite{parameswaran2013fractional, liu2013review,Titusreview,Spanton62}. As we demonstrate in Appendix~\ref{sec:app_fci}, a geometric quench that excites a spin-2 mode can be straightforwardly implemented in an FCI, resulting in the same type of dynamics described by the bimetric theory in the long-wavelength limit. The tunability of parameters in FCIs allows one to observe time scales on the order of $\sim 10^{-5}$s (assuming the interaction strength in cold-atom settings is in the order of $2\pi\hbar\times 10$kHz), which is well within the limits of recent experiments~\cite{Bernien2017}.

New dynamical phenomena will likely arise in Abelian FQH states with more complicated internal structure, such as bi-layer states and Jain states. These states host multiple collective modes which would all contribute to the dynamics. Multi-layer systems feature rich phase diagrams that can be explored by tuning the interaction strength between the layers~\cite{GirvinMacDonaldReview,EisensteinReview}. It would be interesting to map out the dynamical counterparts of such equilibrium phase diagrams by studying the geometric quench in different interaction regimes. 

Even in the simplest models of Laughlin states, our study has focused on the linear regime of adiabatic quench. The non-linear regime should become relevant for strong quenches or quenches in the vicinity of the nematic phase transition. The present quench protocol could be used as a complementary tool for probing the nematic transition, which has recently been studied in Ref.~\cite{regnault2017evidence}. 

When a  ``higher-spin'' pseudopotential drives the quench, the system undergoes non-trivial dynamics which cannot be described within the bimetric theory or by anisotropic trial states \cite{QiuPhysRevB.85.115308}. In these cases, a richer set of trial states is needed to fully describe the dynamics. This set of states should be parametrized by higher-spin cousins of the intrinsic metric $\hat g$. Similar conclusions were reached in Ref.~\cite{GromovSon} (see also Ref.~\cite{Cappelli2016}) where the bimetric theory was argued to be a low-energy approximation to a putative higher-spin theory, and in Ref.~\cite{nguyen2017fractional} where the linearized bimetric for the Jain series at $\nu = n/(2n+1)$ was derived from the Dirac composite Fermi liquid (CFL) theory at large $n$. In the latter approach the higher-spin modes naturally emerge as collective distortions of the composite Fermi surface. In fact, when the CFL state forms, all of the higher-spin degrees of freedom become gapless, and should be  equally important for the dynamics. In these systems, the geometric quench described in the present manuscript, as well as its ``higher-spin'' cousins, will provide an interesting probe of the collective dynamics of the composite fermion Fermi surface. 

The geometric quench also presents an enticing possibility of probing the non-Abelian fractional quantum Hall states, such as the Moore-Read state~\cite{Moore1991362} -- a candidate for the observed $\nu = 5/2$ plateau~\cite{willett1987observation}. Numerical works \cite{bonderson2011numerical, moller2011neutral} have shown that the Moore-Read state hosts a neutral fermionic mode, in addition to the SMA (bosonic) mode. The neutral fermion mode is expected to have angular momentum $3/2$ and is not present in the bimetric theory. It would be very interesting to design the effective theory and numerical probes of this mode at long wavelengths.

Finally, FCIs exhibit much of the same phenomenology as the continuum FQH states (see the recent reviews Refs.~\cite{parameswaran2013fractional, liu2013review,Titusreview} and references therein). For example, the FCIs also feature the GMP algebra~\cite{2014Scaffidi, Roy2014Chern}, and they naturally correspond to anisotropic FQH states~\cite{harper2014perturbative, GeneralizedPPs}. On the other hand, the FCI states in higher Chern number bands \cite{LiuHighC,WangHighC,SterdyniakHighC} do not have the usual continuum FQH states as their counterparts \cite{WuFCIModel}. Given that the quench driven by anisotropy in hopping has been studied for integer Chern insulators~\cite{LatticeQuench1,LatticeQuench2}, it would be interesting to investigate more systematically the quenches for FCIs, starting from the basic method that we have presented in Appendix~\ref{sec:app_fci}.

\acknowledgements

We thank V. Caudrelier for illuminating discussions about the system of differential equations. This work was supported in part by the Department of Energy, Office of  Basic Energy Sciences through Grant No.~DE-SC0002140. Z.L. is supported by the National Thousand-Young-Talents Program of China. Z.L. was additionally supported by Alexander von Humboldt Research Fellowship for Postdoctoral Researchers.
A.G. was supported by the Leo Kadanoff fellowship, the NSF Grant No.~DMR-1206648, and by the University of Chicago Materials  Research  Science  and  Engineering  Center, which is funded by the National Science Foundation under Award No.~DMR-1420709. A.G. was also supported by the Quantum Materials program at LBNL, funded by the U.S. Department of Energy under Contract No.~DE-AC02-05CH11231.  
Z.P. acknowledges support by EPSRC Grants No.~EP/P009409/1 and No.~EP/R020612/1.

Statement of compliance with EPSRC policy framework on research data: This publication is theoretical work that does not require supporting research data.

\appendix

\section{Anisotropic-to-anisotropic quench}\label{sec:anisotropic}

As an immediate generalization of the isotropic-to-anisotropic quench studied in Sec.~\ref{sec:isotropicquench}, we now discuss the more general case of a quench with $A_1>A_0>0$, for which the initial state is already anisotropic. In this case, a simple analytical solution does not exist under the general initial conditions $Q(0)>0,\phi(0)=0$ even for the linearized systems \eqref{eq:phidotweak} and \eqref{eq:Qdotweak}. 
However, by numerically solving the coupled differential equations, we find that the bimetric theory still accurately describes the short-time dynamics of $Q$ and $\phi$, and also captures the oscillatory dynamics very well up to moderate times. 

\begin{figure}[htb]
\includegraphics[width=\linewidth]{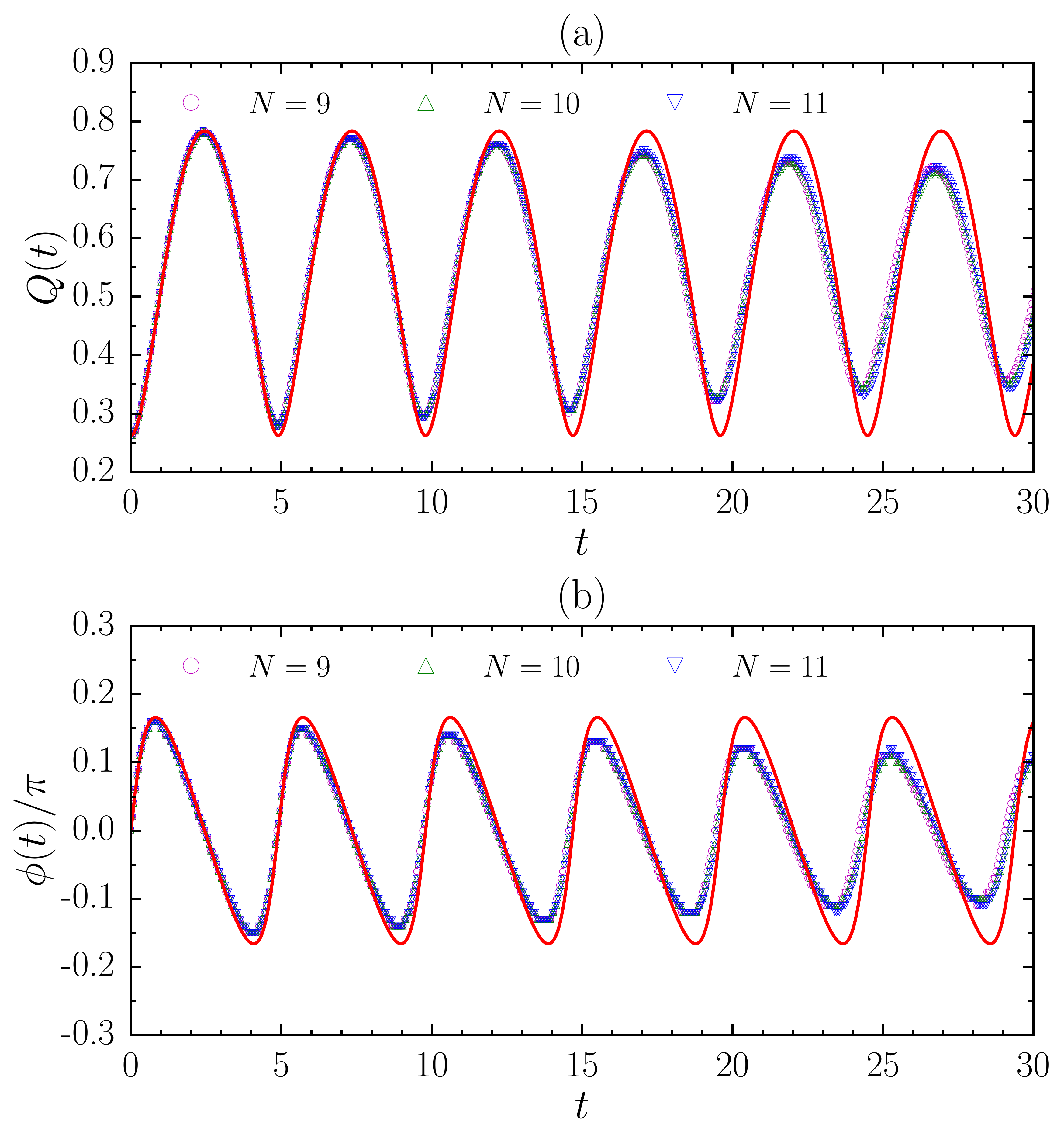}
\caption{Exact dynamics of (a) $Q$ and (b) $\phi$ for $N=9-11$ bosons at $\nu=1/2$ with the contact interaction. The adiabatic geometric quench is driven by choosing $A_0=\ln1.3$ and $A_1=\ln 1.7$. The red solid curve is the solution of linearized equations \eqref{eq:phidotweak} and \eqref{eq:Qdotweak} under the initial condition $Q(0)=A_0,\phi(0)=0$, with $A\approx 0.523$ and $E_\gamma\approx 1.283$. These two parameters are obtained by fitting the first oscillation of $Q$ of $N=11$ against the solution.}
\label{fig:v0_boson2}
\end{figure}
In Fig.~\ref{fig:v0_boson2}, we choose $A_0=\ln 1.3$ and $A_1=\ln 1.7$, and show the exact dynamics for the quench driven by the contact interaction. Since the initial state is a Laughlin model state with $\hat g(0)=g_m$, the bimetric theory predicts the post-quench dynamics of $\hat g$ is governed by the solution of Eqs.~\eqref{eq:phidotweak} and \eqref{eq:Qdotweak} with initial condition $Q(0)=A_0,\phi(0)=0$. Indeed, the numerical data in the first two periods again lie on top of such a solution with fitting parameters $A\approx 0.523$ and $E_\gamma\approx 1.283$. The value of $A$ is close to $A_1=\ln 1.7\approx 0.531$, as expected, and $E_\gamma\approx 1.283$ is consistent with the graviton gap estimated from both the $I_{0,2}(\omega)$ spectral function and the previous fit in Fig.~\ref{fig:v0_boson}. 

As discussed in Sec.~\ref{sec:isotropicquench}, apart from generally good agreement, we can also observe some discrepancy between the exact dynamics and the equations of the bimetric theory in Fig.~\ref{fig:v0_boson2}. This discrepancy is likely caused by the mentioned finite-size effect (the splitting of the spin-$2$ mode into several states and the Lieb-Robinson light cone) and the possible contribution of higher-spin modes. Compared to isotropic-to-anisotropic quench, stronger finite-size effects in the present case are not surprising because anisotropy effectively reduces the length of the system along one of the spatial directions. 

\section{Projected dynamics}\label{sec:projected}

In the main text, we have shown that the number of spin-$2$ eigenstates of $H'$ is exponentially smaller than the full Hilbert space dimension. The number of spin-2 eigenstates was determined by the matrix element of $\hat V_{0,2}$ generalized pseudopotential. Here we investigate how accurately the quench dynamics in Sec.~\ref{sec:isotropicquench} can be reproduced by projecting to this set of states.

\begin{figure}[htb]
\includegraphics[width=\linewidth]{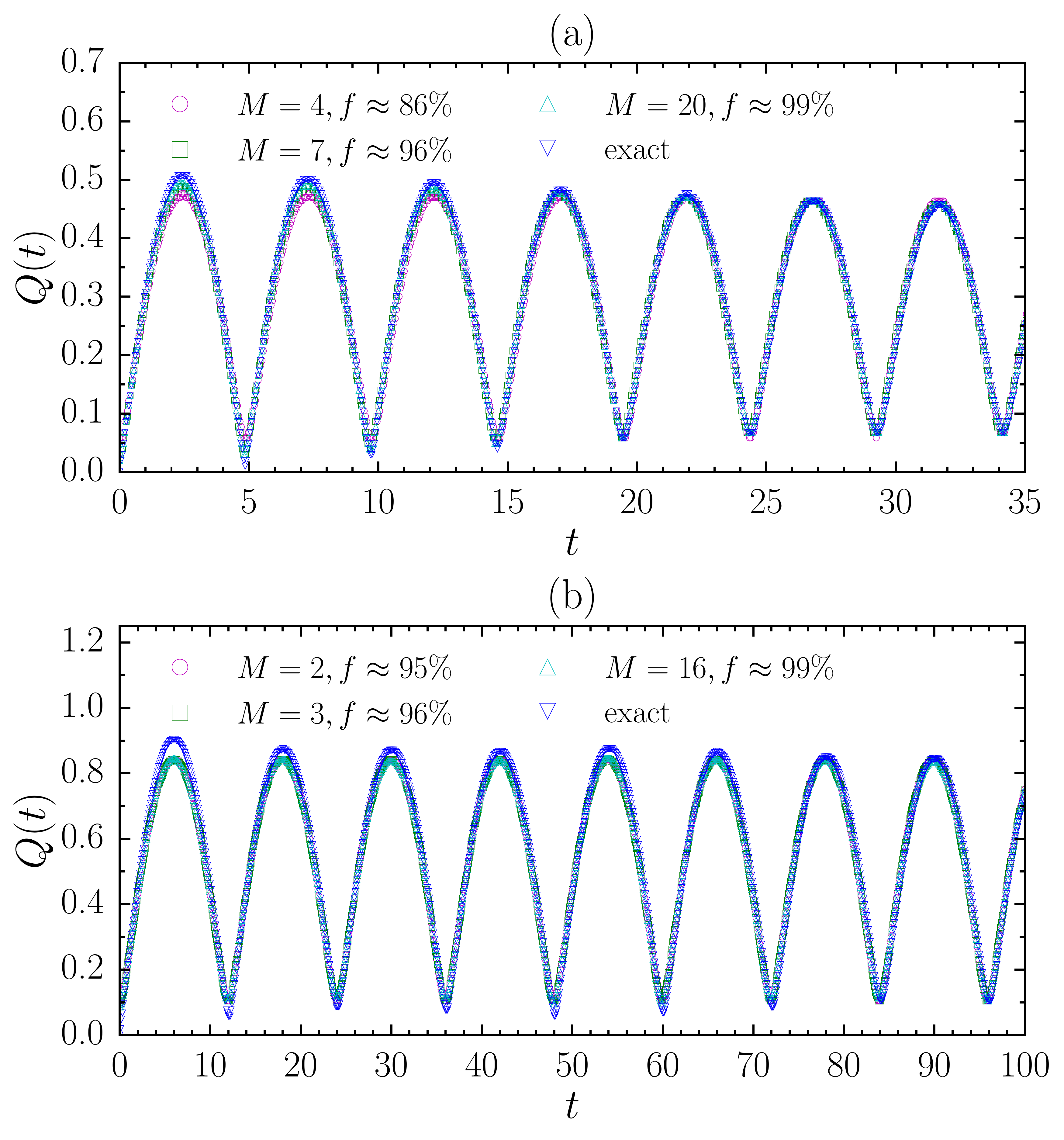}
\caption{Dynamics of $Q$ projected to $M$ eigenstates of $H'$ with dominant $|\langle n| \hat V_{0,2} |0\rangle|^2$. We consider $N=11$ bosons at $\nu=1/2$ with (a) contact and (b) Coulomb interactions. The quench parameters $A_0$ and $A_1$ are the same as those in Figs.~\ref{fig:v0_boson} and \ref{fig:coulomb_boson}, respectively.
}
\label{fig:project_boson}
\end{figure}

We perform the projection explicitly, i.e., we calculate the post-quench state as $|\tilde{\psi}(t)\rangle = \frac{1}{\mathcal{N}}\sum_{n\in \mathcal{S}} \exp(-i \epsilon_n t)\langle n| \psi_0\rangle |n\rangle$, where $\mathcal{S}$ contains the first $M$ eigenstates with the largest $|\langle n| \hat V_{0,2} |0\rangle|^2$ matrix elements and $\mathcal{N}$ is the normalization factor. The dynamics of the intrinsic metric of $|\tilde{\psi}(t)\rangle$ after an isotropic-anisotropic quench is shown in Fig.~\ref{fig:project_boson} for $N=11$ bosons, together with exact dynamics without the projection. As expected, the projected dynamics approaches the exact dynamics with increasing $M$. However, for both contact and Coulomb interactions, the main feature in exact dynamics has been accurately reproduced by $M$ less than $10$, even if the whole Hilbert space in this case is more than four orders of magnitude larger. Such an excellent approximation of the exact dynamics by a tiny spin-$2$ fraction of the whole Hilbert space is an evidence for the picture of graviton oscillation and the bimetric theory.

\section{Non-instantaneous quench} \label{sec:noninst}

\begin{figure}[htb]
\includegraphics[width=\linewidth]{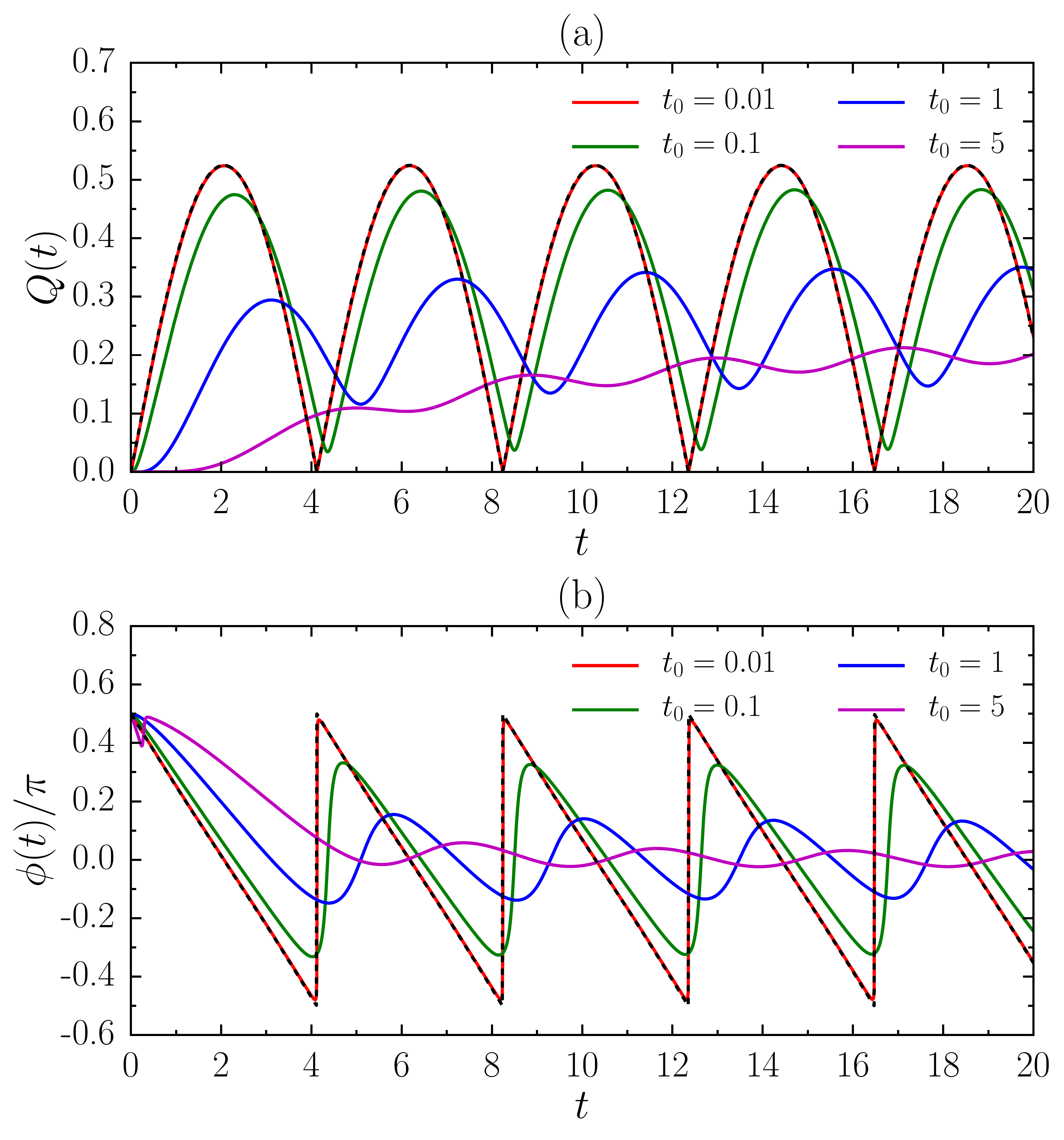}
\caption{Solutions of the full equations [Eqs.~(\ref{eq:phi}) and (\ref{eq:Q})] for the non-instantaneous quench with $A(t)=A_1 \exp{(-t_0/t)}$, under the initial condition $Q(0)=0$. Here we focus on the $Q\geq0$ solution. We choose $\Omega=1.5/4, \gamma=-1, A_1=\ln 1.3$, and consider several different $t_0=0.01$, $0.1$, $1$ and $5$. For comparison, we also give solutions for the instantaneous quench as black dashed lines.}
\label{fig:solution_ramp}
\end{figure}

Instantaneous quench is the minimal theoretical model that describes sudden tilt of the magnetic field. However, since any experimental manipulation takes finite time, we now generalize our discussions to more realistic, non-instantaneous quenches. To be specific, we consider isotropic-anisotropic quenches driven by a time-dependent ramp of the mass tensor, i.e., $g_m={\rm diag}(e^{A(t)},e^{-A(t)})$ with 
\begin{eqnarray}
A(t)=A_1\exp(-t_0/t),
\end{eqnarray}
such that $A(0)=0$ and $A(+\infty)=A_1$. The slope of the ramp depends on some characteristic time, $t_0$. When $t_0=0$, we recover the instantaneous quench studied in Sec.~\ref{sec:isotropicquench}. We thus expect that differences between the non-instantaneous quench and the instantaneous one are negligible when $t_0$ is small, while some qualitative changes may happen for large enough $t_0$. 

As a representative example, we choose $A_1=\ln1.3$ (adiabatic quench) and show the predicted behavior of $Q$ and $\phi$ in the bimetric theory, i.e., the solutions of the full equations [Eqs.~(\ref{eq:phi}) and (\ref{eq:Q})], for different $t_0$ in Fig.~\ref{fig:solution_ramp}. We indeed find that the solutions in the non-instantaneous case agree with their instantaneous limit up to $t_0\lesssim 10^{-2}$. The effect of the ramp starts to appear when $t_0\sim 10^{-1}$, which leads to gradual deviations of both the amplitude and frequency in the dynamics from the instantaneous case. Finally, for very large $t_0\gtrsim 1$, we observe strongly distorted oscillation patterns in $Q$ and $\phi$, revealing very different dynamics between the fast-quench and slow-quench regimes. 

\begin{figure}[htb]
\includegraphics[width=\linewidth]{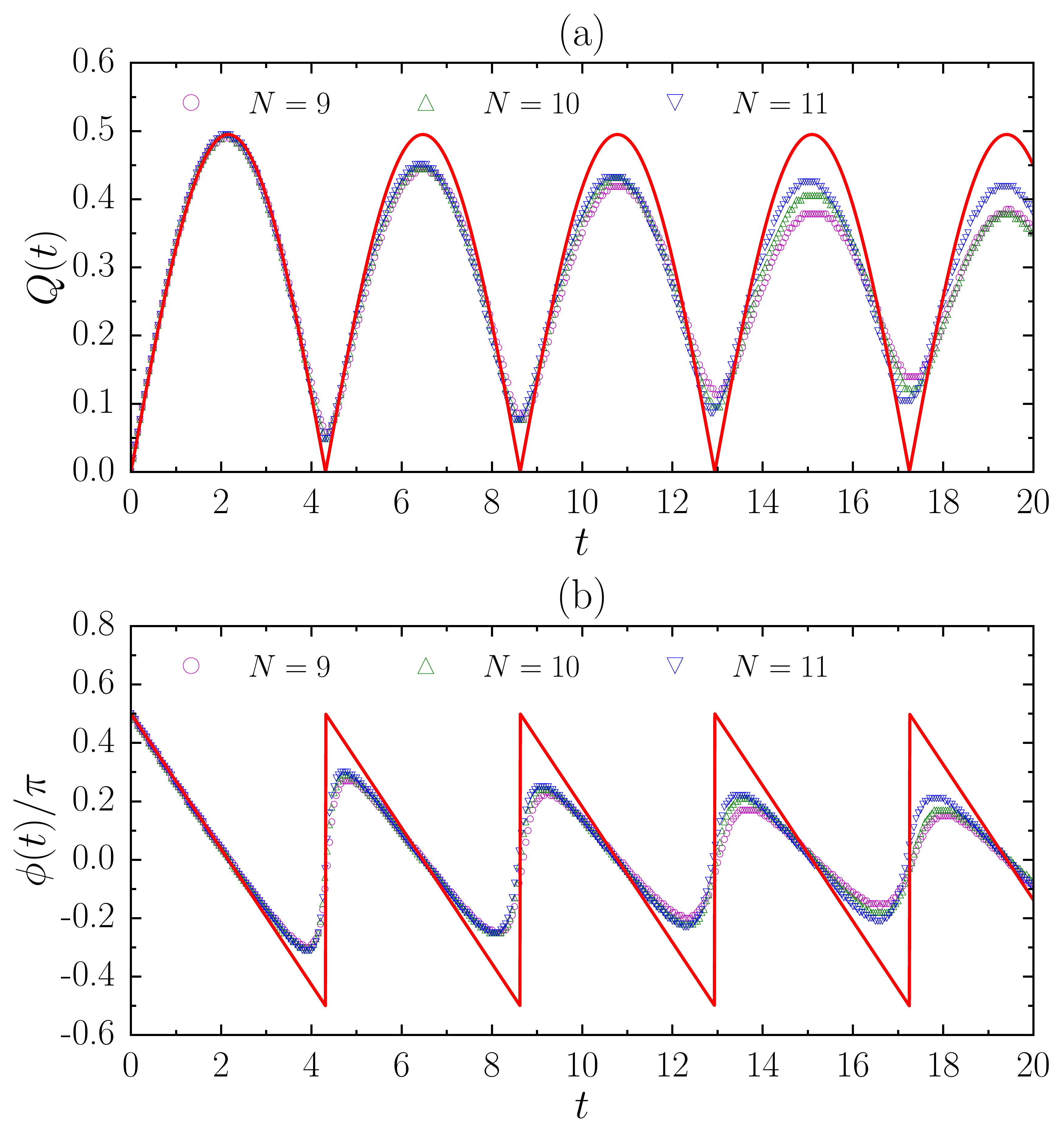}
\caption{Exact dynamics of (a) $Q$ and (b) $\phi$ for $N=9-11$ fermions at $\nu=1/3$ with the $V_1$ interaction. The adiabatic geometric quench is driven by choosing $A_0=0$ and $A_1=\ln 1.3$. The red solid curve is the $Q\geq0$ part of Eqs.~(\ref{eq:linsol1}) and (\ref{eq:linsol2}), with $A\approx 0.248$ and $E_\gamma\approx 1.457$. These two parameters are obtained by fitting the first oscillation of $Q$ of $N=11$ against the solution.}
\label{fig:v1_fermion}
\end{figure}

It would be interesting to compare the exact dynamics of finite systems  with the predictions of the bimetric theory also for non-instantaneous quenches, as we have done for the instantaneous case in Sec.~\ref{sec:results}. Due to the time-dependent nature of the non-instantaneous quench, the many-body numerical simulation in this case is more complicated and will be presented in future work. However, because the instantaneous quench corresponds to a stronger perturbation of the system, we expect similar or better agreement between theory and numerics for the case of non-instantaneous quenches.

\begin{figure}[htb]
\includegraphics[width=\linewidth]{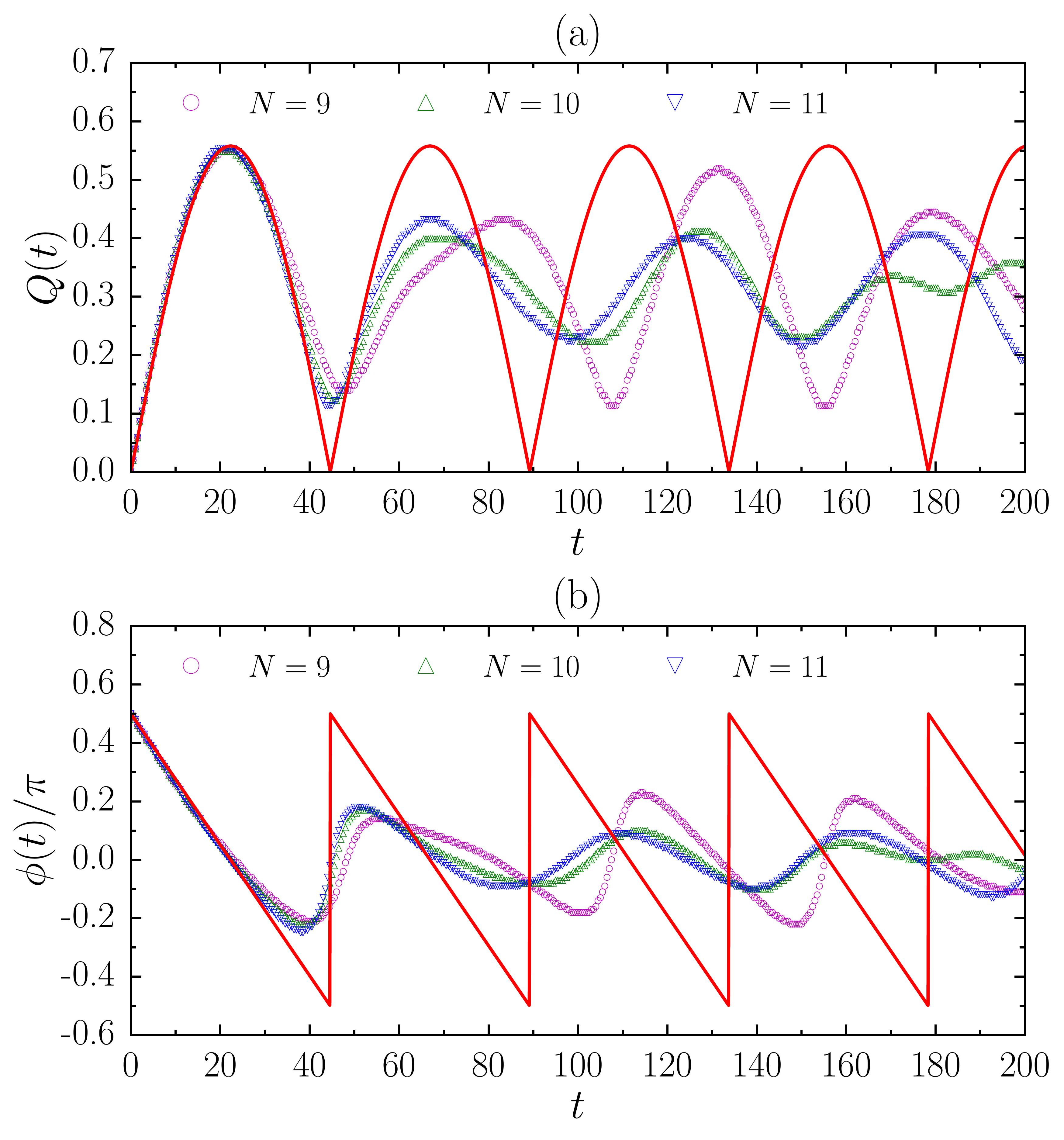}
\caption{Exact dynamics of (a) $Q$ and (b) $\phi$ for $N=9-11$ fermions at $\nu=1/3$ with the Coulomb interaction. The adiabatic geometric quench is driven by choosing $A_0=0$ and $A_1=\ln 2.0$. The red solid curve is the $Q\geq0$ part of Eqs.~(\ref{eq:linsol1}) and (\ref{eq:linsol2}), with $A\approx 0.279$ and $E_\gamma\approx 0.141$. These two parameters are obtained by fitting the first oscillation of $Q$ of $N=11$ against the solution.}
\label{fig:coulomb_fermion}
\end{figure}

\section{Adiabatic geometric quench for $\nu=1/3$ fermions}\label{sec:app_fermions}

Here we show the numerical results of the exact dynamics of $\hat g$ after a mass anisotropy quench for $\nu=1/3$ fermions. As representative examples, we focus on adiabatic isotropic-anisotropic quenches with $A_0=0$ and $A_1>0$, where we choose $A_1=\ln1.3$ and $A_1=\ln2.0$ for quenches driven by $V_1$ and Coulomb interactions, respectively. Exact dynamics of $\hat g$ up to moderate times is shown in Figs.~\ref{fig:v1_fermion} and \ref{fig:coulomb_fermion} for these two types of interactions.

Similar to $\nu=1/2$ bosons, the short-time dynamics of adiabatic quenches for $\nu=1/3$ fermions fully agrees with the bimetric theory. We can thus extract parameters $A$ and $E_\gamma$ by fitting the numerical data against Eqs.~(\ref{eq:linsol1}) and (\ref{eq:linsol2}), whose values are given in the captions of Figs.~\ref{fig:v1_fermion} and \ref{fig:coulomb_fermion}. The value of $A$, which quantifies the anisotropy in the interaction, is again close to $A_1$ (between $0$ and $A_1$) for the $V_1$ (Coulomb) interaction as expected. The value of $E_\gamma$ estimates the graviton gap as $1.46$ and $0.14$ for $V_1$ and Coulomb interactions, respectively. Note that the graviton gap of the Coulomb interaction that we extract from the quench dynamics precisely matches the value given by other methods in the context of single-mode approximation \cite{GMP85,GMP86}.

Like in the bosonic case, we also observe deviations from the bimetric theory at longer times for $\nu=1/3$ fermions. However, these deviations are more serious and appear earlier than those for $\nu=1/2$ bosons. For quenches driven by the $V_1$ interaction, while the overall oscillation structure is maintained up to moderate times, the oscillating amplitude already visibly decays after the first oscillation (Fig.~\ref{fig:v1_fermion}). The situation is even worse for quenches driven by the Coulomb interaction, as expected. In that case, the behavior of $Q$ and $\phi$ totally departs from simple oscillations after the first period and the curves of different system sizes are no longer on top of each other (Fig.~\ref{fig:coulomb_fermion}). Compared with $\nu=1/2$ bosons, our numerical results clearly demonstrate that the dynamics of $\nu=1/3$ fermions suffers more from finite-size effects in the excitations and the participation of higher-spin modes.

\section{Fractional Chern insulators}\label{sec:app_fci}

Here we generalize the geometric quench from continuum FQH states to their lattice analogs -- the fractional Chern insulators (FCIs) \cite{parameswaran2013fractional, liu2013review,Titusreview}. Contrary to conventional FQH states, FCIs do not require an external magnetic field~\cite{sun2011nearly,neupert2011fractional} and may potentially persist at higher temperatures~\cite{tang2011hightemperature}. 
Considering the recent experimental progress on quantum many-body dynamics in optical lattices \cite{opticallatticequench}, FCIs realized by cold atoms in optical lattices could be a promising platform to observe the dynamics after a geometric quench on much longer time scales than possible in semiconductor FQH systems. In the following, we will describe a quench protocol for FCIs and demonstrate that our main results directly apply to this type of system.
As FCIs are inherently anisotropic, the quenches in FCI belong to the anisotropic-anisotropic case discussed in Appendix~\ref{sec:anisotropic}.

For concreteness, we consider $N$ bosons in a two-dimensional Ruby lattice \cite{rubylattice} of $N_1\times N_2$ unit cells. The system is in $xy$ plane with periodic boundary conditions. We adopt the same tight-binding parameters and lattice configurations as in Ref.~\cite{wu2012zoology}, for which the lowest Bloch band is flat and has Chern number $C=1$. We assume that bosons interact via onsite and dipolar potentials. If all dipoles are polarized in $xz$ plane with an angle $\alpha$ to the $+z$ direction, the interaction Hamiltonian can be written as 
\begin{eqnarray}
H_{\rm int}=\sum_{i}n_i (n_i-1)+\sum_{i<j}\frac{r_{ij}^2-3x_{ij}^2\sin^2\alpha}{r_{ij}^5}n_i n_j,\nonumber\\
\label{fciint}
\end{eqnarray}
where $n_i$ is the occupation on lattice site $i$, and $r_{ij}=(x_{ij},y_{ij})$ is the distance between lattice site $i$ and $j$. We project the interaction to the lowest band and verify that the ground state at band filling $\nu=N/(N_1N_2)=1/2$ is the $\nu=1/2$ bosonic Laughlin FCI for a finite range of $\alpha$, characterized by a robust two-fold topological degeneracy. 
In order to keep the lattice aspect ratio close to $1$ and for the two degenerate FCI states to be located in different momentum sectors, we focus on two system sizes below: $N=6,N_1=3,N_2=4$ and $N=10,N_1=4,N_2=5$. 

\begin{figure}[htb]
\includegraphics[width=\linewidth]{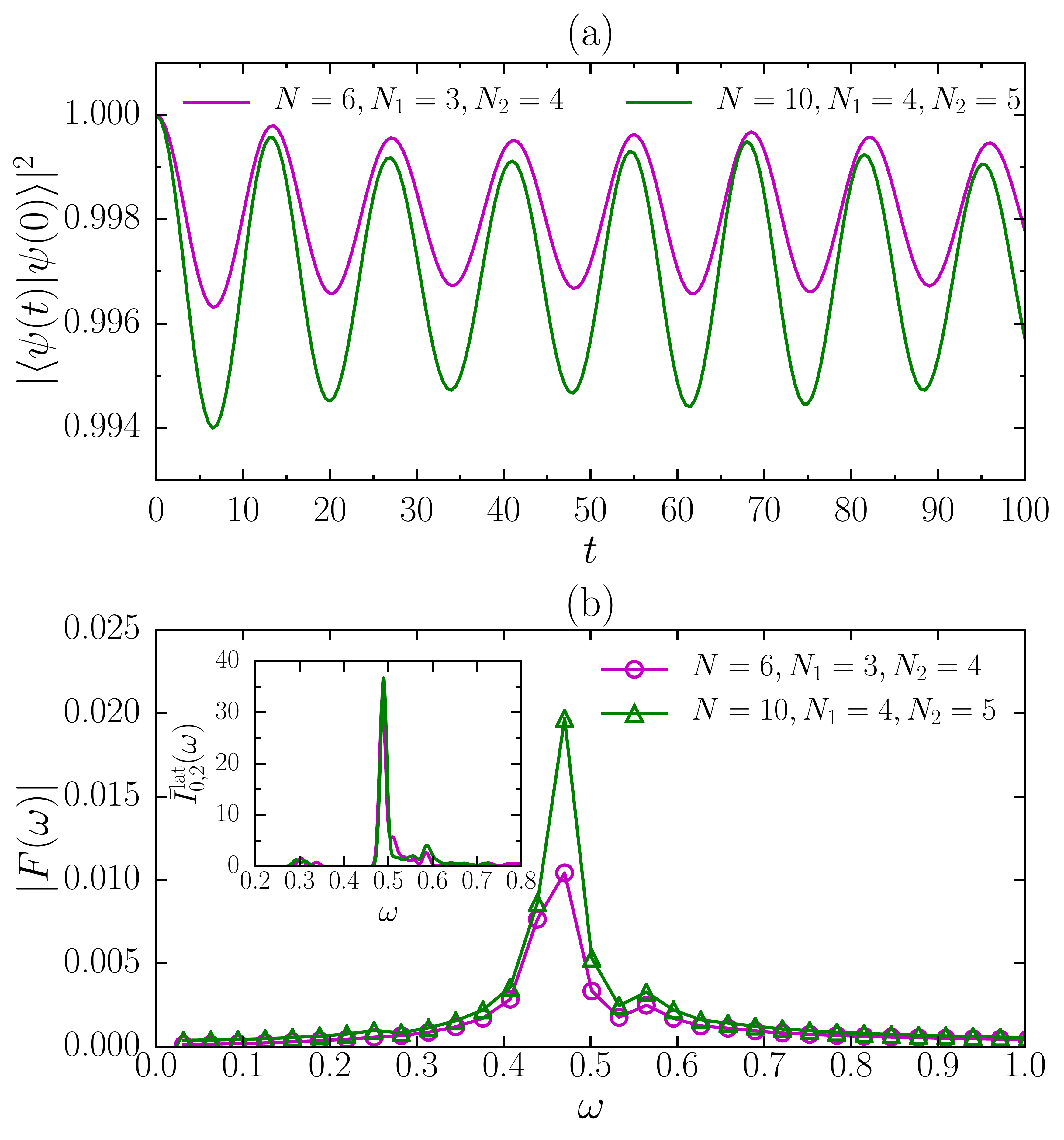}
\caption{(a) Dynamics of fidelity, $|\langle\psi(t)|\psi(0)\rangle|^2$, and (b) its discrete Fourier transform for $\nu=1/2$ bosonic Laughlin FCIs. The adiabatic quench is driven by $\alpha'=15^{\circ}$. The inset of (b) shows the normalized spectral function ${\bar I}_{0,2}^{\rm lat}(\omega)=I_{0,2}^{\rm lat}(\omega)/\int I_{0,2}^{\rm lat}(\omega) d\omega$ for the isotropic case with $\alpha=0$. 
Markers in insets and curves in main figures with the same color refer to the same system size.}
\label{fig:fci}
\end{figure}

While the interaction is isotropic for $\alpha=0$, tuning $\alpha$ away from $0$ adds anisotropy, which can be experimentally achieved by changing the direction of the polarizing magnetic field \cite{Baier201}. We prepare the initial state as the ground state with momentum ${\bf K}={\bf 0}$, then drive the quench by suddenly changing $\alpha$ from $0$ to a small but finite $\alpha'$ (adiabatic quench). Tuning $\alpha$ is not a direct lattice analog of changing the geometry of continuum Landau level orbitals, but it is also expected to excite a geometric degree of freedom in FCIs because it adds quadrupolar anistropy in the system. In order to confirm this, we track the dynamics of fidelity $|\langle\psi(t)|\psi(0)\rangle|^2$, and compare its dominant frequency with the energy of FCI graviton mode which can be probed by a suitable lattice analog of the $\hat{V}_{0,2}$ pseudopotential. A detailed study of intrinsic metric is also feasible for FCIs by the FCI-FQH mapping \cite{WuFCIModel} and will be presented in a future work. 

In Fig.~\ref{fig:fci}, we show the fidelity and its discrete Fourier transform for $\alpha'=15^{\circ}$. One can see that the dynamics of fidelity is dominated by a single frequency, characterized by the pronounced peak at $\omega\approx 0.47$ in the discrete Fourier transform [Fig.~\ref{fig:fci}(b)]. In order to estimate the graviton energy on the lattice, we define the operator
\begin{eqnarray}
\hat{V}_{0,2}^{\rm lat}=\sum_{{\bf q}}(\cos q_1-\cos q_2){\bar{\rho}_{\bf q}}^{\rm lat} \bar{\rho}_{-{\bf q}}^{\rm lat},
\end{eqnarray}
where $q_i={\bf q}\cdot{\bf b}_i$ with ${\bf b}_i$'s the two basic lattice direction vectors, ${\bf q}$ is in the first Brillouin zone, and $\bar{\rho}_{\bf q}^{\rm lat}$ is the density operator $e^{i{\bf q}\cdot\hat{{\bf r}}}$ projected to the lowest Bloch band (see Refs.~\cite{Sondhidensity,Nicolasdensity,WuFCIModel,Repellin} for the explicit form of $\bar{\rho}_{\bf q}^{\rm lat}$). Similar to the continuum $\hat{V}_{0,2}$ pseudopotential, $\hat{V}_{0,2}^{\rm lat}$ effectively carries angular momentum $L_z=2$ to probe spin-$2$ states on the lattice, because the ${\bf q}$-dependent term $\cos q_1-\cos q_2$ takes a $d$-wave form $q_1^2-q_2^2$ at small $|{\bf q}|$.
Indeed, its spectral function
\begin{eqnarray}
I_{0,2}^{\rm lat}(\omega) = \sum_j \delta(\omega-\epsilon_j) |\langle j | \hat{V}_{0,2}^{\rm lat} |0\rangle|^2
\end{eqnarray} 
shows sharply pronounced peaks [inset of Fig.~\ref{fig:fci}(b)],
where $\epsilon_j$ and $|j\rangle$ are the eigen-energy and eigenstate of the projected interaction (\ref{fciint}). Since these peaks appear around the same energy as that in the discrete Fourier transform of fidelity, our quench protocol indeed excites the geometric degree of freedom of FCI that corresponds to the spin-$2$ graviton mode on a lattice. The similarity of the FCI results to the continuum FQH is rather surprising, in light of the fact that there are no quantitative arguments that support the relation between the bimetric theory and FCIs (beyond the intuition mentioned in Ref.~\cite{GromovSon}).

\bibliography{fqhe}

\end{document}